\documentclass[12pt]{article}
\usepackage{scicite}
\usepackage{times}
\usepackage{bm}
\usepackage{amsmath}
\usepackage{amssymb}
\usepackage{graphicx}
\pdfpageattr {/Group << /S /Transparency /I true /CS /DeviceRGB>>}

\newenvironment{sciabstract}{%
\begin{quote} \bf}
{\end{quote}}

\newcounter{lastnote}

\title{Visualization of superparamagnetic dynamics in magnetic topological insulators}

\author
{E. Lachman$^{1,\ast}$, A. F. Young$^{1,2,\ast}$, A. Richardella$^3$, J. Cuppens$^1$, Naren HR$^1$,\\ Y. Anahory$^1$, A. Y. Meltzer$^1$,  A. Kandala$^3$, S. Kempinger$^3$,\\ Y. Myasoedov$^1$, M. E. Huber$^4$, N. Samarth$^3$ and E. Zeldov$^{1}$ \\
\normalsize{$^{1}$Department of Condensed Matter Physics, Weizmann Institute of Science}\\
\normalsize{Rehovot 76100, Israel}\\
\normalsize{$^{2}$Physics Department, University of California}\\
\normalsize{Santa Barbara 93106-9530, USA}\\
\normalsize{$^{3}$Department of Physics, The Pennsylvania State University}\\
\normalsize{University Park, Pennsylvania 16802, USA}\\
\normalsize{$^{4}$Department of Physics, University of Colorado Denver}\\
\normalsize{Denver, Colorado 80217, USA}\\
\normalsize{$^{*}$These authors contributed equally}\\
\normalsize{e-mail:eli.zeldov@weizmann.ac.il}
}

\date{}

\begin{document}

\baselineskip24pt
\maketitle

\begin{sciabstract}
Quantized Hall conductance is a generic feature of two dimensional electronic systems with broken time reversal symmetry. In the quantum anomalous Hall state recently discovered in magnetic topological insulators, time reversal symmetry is believed to be broken by long-range ferromagnetic order, with quantized resistance observed even at zero external magnetic field. Here, we use scanning nanoSQUID magnetic imaging to provide a direct visualization of the dynamics of the quantum phase transition between the two anomalous Hall plateaus in a Cr-doped (Bi,Sb)$_2$Te$_3$ thin film. Contrary to naive expectations based upon macroscopic magnetometry, our measurements reveal a superparamagnetic state formed by weakly interacting magnetic domains with a characteristic size of few tens of nanometers. The magnetic phase transition occurs through random reversals of these local moments, which drive the electronic Hall plateau transition. Surprisingly, we find that the electronic system can in turn drive the dynamics of the magnetic system, revealing a subtle interplay between the two coupled quantum phase transitions.
\end{sciabstract}

The integer quantum Hall effect (QHE), first observed in clean two dimensional electron systems at high magnetic fields \cite{Klitzing1980}, is the paradigmatic example of a topological phase: different integer quantum Hall states are characterized by identical symmetries but different integer topological quantum numbers $\eta$, with the quantized Hall resistance given by $R_{xy}=\eta h/e^2$ \cite{Thouless1982}. However, Hall quantization may occur also in the absence of an external field as long as time-reversal symmetry (TRS) is broken \cite{Haldane1988}. This quantum anomalous Hall (QAH) state was recently realized experimentally \cite{Chang2013, Checkelsky2014, Kou2014, Bestwick2015, Chang2015, Kandala2015} following theoretical proposals based on combining strong spin-orbit coupling with long-range ferromagnetic (FM) order in magnetically doped topological insulators (TI) \cite{Hasan2010, Qi2011, Onoda2003, Liu2008, Qiao2010, Yu2010, Ezawa2012, Garrity2013, Xu2014}. The QH plateau transition represents the canonical example of a topological phase transition, described by the divergence of the localization length and a universal critical scaling of transport coefficients\cite{Huckestein1995}. Recent theoretical calculations show that under certain assumptions, the QAH plateau transition can be mapped onto the same network model used to describe the integer QH plateau transition, leading to the same scaling laws\cite{Wang2014}.

While the electronic topological transition in QH and QAH systems, taken in isolation, can be viewed as identical, the two experimental systems differ in several key aspects. The QH plateau transition is a purely electronic effect, in which delocalization occurs against a background of quenched electronic disorder. In contrast, the QAH plateau transition results from two coupled quantum phase transitions: the field-driven magnetic transition of the FM order and the electronic transition that is driven by TRS breaking of the FM transition. The dynamics of the FM reversal can endow the QAH electronic transition with features that do not have an analog in QHE systems. Exactly at the transition, both systems can be understood as a network of domain walls---magnetic domain walls for the QAH, and domain walls of different filling factor in the QHE---that host counterpropagating chiral edge states. Crucially, however, in contrast to the QH plateau transition, which describes the phase diagram of the system in thermodynamic equilibrium, the FM domain structure in the QAH is metastable leading to hysteresis and relaxation dynamics that can directly affect the electronic system. Moreover, the energy scale of the electronic delocalization transition and its critical scaling may depend on the details of the microscopic FM structure and on the scaling of the magnetic phase transition. Finally, the electronic system, which commonly mediates FM interactions in dilute magnetic systems, can in turn modify the phase transition of the magnetic system. Most previous studies of QAH systems have used electronic transport measurements to probe the combined effect of magnetic and electronic evolutions, making it difficult to disentangle the individual roles of the two phase transitions. Exploring the reciprocal coupling of the magnetic and electronic quantum phase transitions thus requires selective measurement tools that can address the two systems independently.

Here we combine electronic transport with a scanning superconducting quantum interference device (SQUID) of 200 nm diameter that resides on the apex of a quartz tip (SQUID-on-Tip (SOT)) \cite{Finkler2010,Vasyukov2013} to simultaneously probe the magnetic and electronic transitions at $^3$He temperatures in a 7 quintuple layer (QL) thick Cr$_{0.1}$(Bi$_{0.5}$Sb$_{0.5}$)$_{1.9}$Te$_3$ film grown by molecular beam epitaxy on a SrTiO$_3$ substrate (Fig. 1). The large dielectric constant of the substrate allows effective control of the chemical potential through back gating. The hysteretic transition between two Hall resistance ($R_{xy}$) plateaus occurs at a coercive field of $H_c=130$ mT at 250 mK (Fig. 1A). At elevated fields, the longitudinal resistance ($R_{xx}$) shows a pronounced dip as a function of the back gate voltage at $V_{g}\approx6$ V due to an incipient QAH state (Fig. 1B). At the lowest measurement temperature (250 mK) used here, the sample we discuss does not yet show the fully developed QAH. This is similar to previous experiments on Cr-doped topological insulator films\cite{Chang2013, Checkelsky2014, Kou2014, Bestwick2015, Chang2015, Kandala2015} where quantized Hall resistance appears only at dilution refrigerator temperatures, far below the onset temperature of hysteretic magnetic behavior\cite{Chang2013, Chang2013a}. Figures 1E-H show images of the local distribution of the magnetic field $B_z(x,y)$ in the sample at various points along the magnetization loop. We find $B_z(x,y)$ to be highly inhomogeneous at all fields, with peak contrast at $H_c$, where the average magnetization vanishes (Figs. 1F,G). Surprisingly, submicron structure is evident even at fields corresponding to saturation of transport coefficients (Figs. 1E,H). Images corresponding to opposite magnetization are highly anticorrelated on microscopic scales---including at full saturation. This suggests an inhomogeneous distribution of magnetic moments due to segregation of the Cr dopants. While high resolution transmission electron microscopy measurements on our samples have yet to show any obvious evidence of Cr clustering (see Fig. S11), we cannot preclude inhomogeneity at the nanoscale, akin to that found in other magnetically doped semiconductors such as Cr-doped ZnTe \cite{Kuroda2007}. Clear evidence of phase separation has only been seen in Cr doped Bi$_2$Se$_3$ thin film ~\cite{Chang2014} samples that do not show the QAH state, while large nanoscale fluctuations in the local Cr density have been observed in Te based samples ~\cite{Lee2015} due to random doping that introduces strong disorder in the material \cite{Beidenkopf2011}.

In metallic FM thin films with out-of-plane magnetization, it is well established that magnetization reversal develops via the nucleation and propagation of domain walls (DWs) separating regions of opposite magnetization. Such DW mediated magnetization reversal has also been imaged in ferromagnetic semiconductor films \cite{Yamanouchi2007,Balk2011} at magnetic dopant concentrations comparable to those in our magnetic TI films. However, scanning SOT microscopy reveals a very different picture of the magnetization reversal process in magnetic TIs. Figure 2A shows a sequence of $B_z(x,y)$ images acquired for increasing values of $\mu_0H$ near $H_c$. The five images appear almost identical; however, numerical subtraction of successive image data ($\Delta B_z(x,y)$, see Fig. 2B) reveals the underlying dynamic process. Instead of the anticipated DW motion, magnetization reversal occurs through a series of random events in which isolated nanoscale islands undergo a reversal of their out-of-plane magnetic moment (see Movie S1). As we discuss below, this constitutes a direct microscopic observation of superparamagnetism in magnetically doped TI films. Our observations caution against drawing conclusions about the ferromagnetic state solely from macroscopic magnetization probes (SQUID magnetometry, magneto-optical Kerr effect) which show square hysteresis loops with robust zero field remanence \cite{Kou2013}.

To quantify the superparamagnetic dynamics across the Hall plateau transition, we fit each of the local features in the $\Delta B_z(x,y)$ maps with a point-like out-of-plane magnetic moment $m$ (Fig. S3).
Figure 2C summarizes $\sim$1700 such fits, accumulated over four different ranges of magnetic field. Throughout the measured range, the spatial distribution of reversal events is random (Fig. 2D inset and Fig. S4), suggesting weak interactions between neighboring islands and supporting the conclusion that superparamagnetism in Cr-doped BiSbTe films is characterized by aggregations of dopant atoms. The deviation of the Hall resistance from its saturation value begins around $\mu_0H=0$ and is accompanied by the reversal of nanoscale moments with average $\bar{m} = 3\times 10^4 \mu_\textrm{B}$ (Figs. 2C,D). Given an average saturation magnetization of $\sim3 \mu_\textrm{B} / \textrm{Cr}$ atom as obtained from global magnetization measurements (Fig. S14), the estimated average diameter of these flipping islands is $d= 51$~nm for our 7~nm thick film, considerably below our spatial resolution of $\sim$300~nm. As the field is increased towards $H_c$, however, a pronounced change in the moment distribution is observed with a shift to higher $m$ values and appearance of a long tail of large moments with $m\gtrsim$ 10$^5\mu_\textrm{B}$ (Figs. 2C,D). The microscopic reversal moments $m$ can be summed to obtain the net change in magnetization $M$ over a continuous field range (Fig. 2E). Comparison with simultaneously acquired $R_{xy}$ shows a qualitative match, implying that the behavior of the transport coefficients through the plateau transition is mainly determined by the underlying change in magnetization. Superparamagnetic behavior and a similar relationship between measured magnetization and transport coefficients were found in a second sample as well as in a Mn-doped Bi$_2$Te$_3$ film (see Movie S2 and Figs. S2, S12-13 and S15).

Hysteretic magnetic transitions are a signature of metastability, and typically display temporal relaxation via thermal activation or quantum tunneling. We probe magnetic relaxation in real time by polarizing the system at $\mu_0H=-1$ T and then ramping the field to a positive set point $\mu_0H_{set}$. We then acquire repeated images of $B_z(x,y)$  while simultaneously monitoring electronic transport. No relaxation is observed on laboratory time scales for $\mu_0H_{set} < 0$ in either magnetization or transport coefficients. Spontaneous relaxation begins to be evident at small positive fields ($\mu_0H_{set}=63$ mT), manifesting as magnetic reversal events $\Delta B_z(x,y)$ and a slow upward drift of $R_{xy}$. The frequency and number of these reversals increases significantly near the coercive field at $\mu_0H_{set}=126$ mT (Figs. 3A,C).

The temporal relaxation measurements further corroborate the superparamagnetic behavior: at temperatures well below the blocking temperature, the magnetization of a superparamagnet is hysteretic, showing minimal relaxation at low fields. On approaching $H_c$ the magnetic anisotropy barrier $U$ is reduced, leading to relaxation when $U\simeq k_\text{B} T$. Since $U$ is proportional to the volume of the superparamagnetic particles, smaller islands undergo thermal activation at a lower field. Simultaneous transport measurements (Fig. 3D) indicate that the electronic transition closely tracks the magnetic relaxation, with transport coefficient relaxation evident at $\mu_0H_{set}=63$ mT and pronounced at 126 mT, in accord with the total temporal change in magnetization extracted from the SOT data (Fig. 3C).

The plateau transition observed in electronic transport appears to be mainly dictated by the underlying magnetic reversal. Surprisingly, however, we find that the dynamics of the magnetic system can in turn be influenced by the electronic system. To explore this effect, we perform a sequence of magnetic imaging at 126 mT interspersing consecutive scans with small excursions of the back gate voltage $\Delta V_{g}$ (Fig. 3B). Remarkably, even small $\Delta V_{g}\sim 1$ V excursions enhance the relaxation of the superparamagnetic islands significantly, as is evident from the statistics of the observed moment reversals (Fig. 3C).  This enhancement in turn is evident in the transport coefficients, as shown in Fig. 3D.

A full comparison of the effects of applied magnetic field, gate voltage, and time on transport coefficients is presented in Fig. 4C, which shows a unified parametric plot of the transport coefficient vector ($R_{xx}$, $R_{xy}$). On this plot, the large magnetic hysteresis evident in Figs. 4A,B is absent, demonstrating that the relation between $R_{xx}$ and $R_{xy}$ (at a given $V_{g}$) is a universal function determined by the magnetization. Within a single constant-$V_g$ arc-shaped plot, zero net magnetization corresponds to the maximum of the arc at zero Hall angle, while the varying Hall angle along the arc reflects the varying sample magnetization. Variable $V_g$ traces over $\Delta V_{g} = \pm 30$ V at $\mu_0H=\pm1$ T, marked in black, show contours of variable carrier density at constant saturated magnetization.  Note that at full saturation, variable $V_g$ data retraces the same path upon repetition.

On this plot, the temporal relaxation of transport coefficients (Fig. S8) is seen to track constant $V_g$ arcs, implying that temporal magnetic relaxation is the dominant mechanism and consistent with the scanning magnetometry results (Figs. 4C,D, gray dots along $V_g=6$ V arc). Again consistent with magnetometry data, $V_g$ excursions enhance the magnetization relaxation dramatically in the metastable regime (Figs. 4 C,D). The blue trace shows the evolution of the transport coefficients during three $V_g$ excursions with $\Delta V_{g}$ increasing from $\pm 1$ V to $\pm 3$ V at a constant field of 126 mT. Successive $V_g$ sweeps do not retrace each other, consistent with an irreversible change in the net magnetization. Repeating the same experiment for larger excursion of $\Delta V_{g} = \pm 30$ V, the magnetization is observed to rapidly relax, nearly reaching complete saturation (cyan in Figs. 4C,D). As Fig. 4C makes clear, the gate-induced magnetic relaxation is strongly dependent on maximum extent of the voltage excursion, although we find it to be independent of sweep direction and rate.

Strikingly, despite the dramatic effect of gate voltage variations on magnetic relaxation, field sweeps at different values of $V_{g}$ show only small deviations in the coercive field (Figs. 4A,B) indicating that carrier density has little effect on the average magnetization. These seemingly contradictory observations can be qualitatively understood by invoking the strongly disordered nature of the superparamagnetic state. In our strongly disordered system, the global gap in the density of states is bridged by a proliferation of subgap states induced by the random polarization of the superparamagnetic islands.  These localized states can in turn mediate the FM, in addition to the proposed global van Vleck mechanism that arises from band states\cite{Yu2010}. $V_g$ excursions may modify the magnetic anisotropy energy of individual islands through the strong dependence of local density of states on position and energy, randomly changing the magnetic potential landscape without significantly changing the average magnetization and coercive field. Unlike in QHE systems, however, this disorder is not quenched: the weakly interacting superparamagnetic islands are metastable.  At any positive field, some local configurations are separated from the ground state by small energy barriers $U$, and any local perturbation can cause irreversible flipping of some of the islands leading to rapid stimulated relaxation upon repeated excursions of $V_{g}$. This mechanism is of course enhanced near $H_c$. Our statistical analysis of the flipping moments at various $V_g$ in Fig. S5 qualitatively supports this mechanism.

The observed superparamagnetic state and its rich dynamics emphasize the intricate coupling between the electronic and magnetic sub-systems in magnetic TIs. Our results suggest that the quantum phase transition between QAH states can be strongly affected by the nature of the underlying magnetic transition leading to deviations from the expected universal scaling of the plateau transition as indicated by recent transport studies \cite{Kou2015}. Even though the topological state is robust to the presence of magnetic disorder at very low temperatures, nanoscale superparamagnetism may be responsible for the fragility of the QAH state at elevated temperatures.

\bibliographystyle{ieeetr}

Acknowledgements\\
We thank E. Berg for illuminating discussions. This work was supported by the Minerva Foundation with funding from the Federal German Ministry of Education and Research and by the Israel Science Foundation (grant No 132/14). NS, AR, AK and SK acknowledge support from DARPA MESO (Grant No. N66001-11-1-4110), ONR (Grant No. N00014-12-1-0117) and ARO MURI (Grant No. W911NF-12-1-0461), as well as use of the NSF National Nanofabrication Users Network Facility at Penn State. AFY acknowledges the support of Goldschleger Center for Nanophysics.

Additional images and data obtained in the present work are presented in the supplementary materials.

\clearpage

%%%%%%%%%%%%%%%% FIGURE 1 %%%%%%%%%%%%%%%%
\begin{figure*}[ht!]
	\begin{center}
\includegraphics[scale=1]{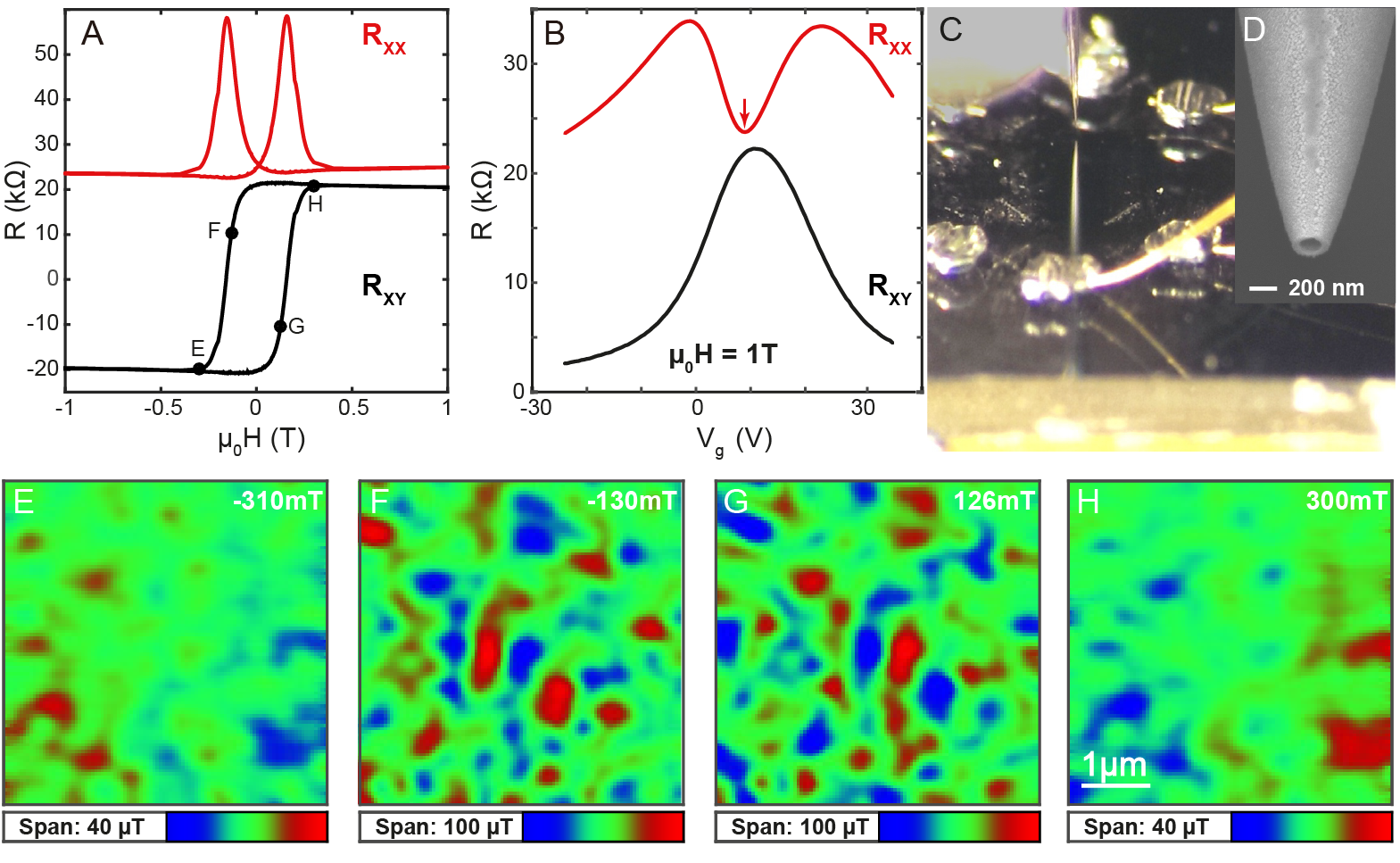}
\caption{\textbf{Electrical transport and scanning magnetic imaging of 7QL thick Cr$_{0.1}$(Bi$_{0.5}$Sb$_{0.5}$)$_{1.9}$Te$_3$ sample at T=250 mK.}
(A-B) Transport measurements showing magnetic field dependence of $R_{xx}$ (red) and $R_{xy}$ (black) at $V_g=6$ V (A) and the $V_g$ dependence at 1 T (B). The dip in $R_{xx}$ marked by the arrow shows the incipient QAH state. (C) An optical image of the sample and SOT showing the electrical contacts and the SOT reflection from the sample surface. (D) Electron micrograph of the SOT used for the magnetic imaging. (E-H) Scanning SOT images $5\times5$ $\mu$m$^2$ of the out-of-plane magnetic field $B_z(x,y)$ at $\sim 300$ nm above the sample surface at four anti-symmetric locations along the magnetization loop marked in (A). Note strong anti-correlation between (E) and (H), and (F) and (G). Pixel size 50 nm, pixel integration time 10 ms.}
	\end{center}
\vspace{0mm}
\end{figure*}	
%%%%%%%%%%%%%%%%%%%%%%%%%
%
%%%%%%%%%% FIGURE 2 %%%%%%%%%%%%%%%%%%%%%%%%%%%
\begin{figure*}[ht!]
	\begin{center}
\includegraphics[width=6.5 in]{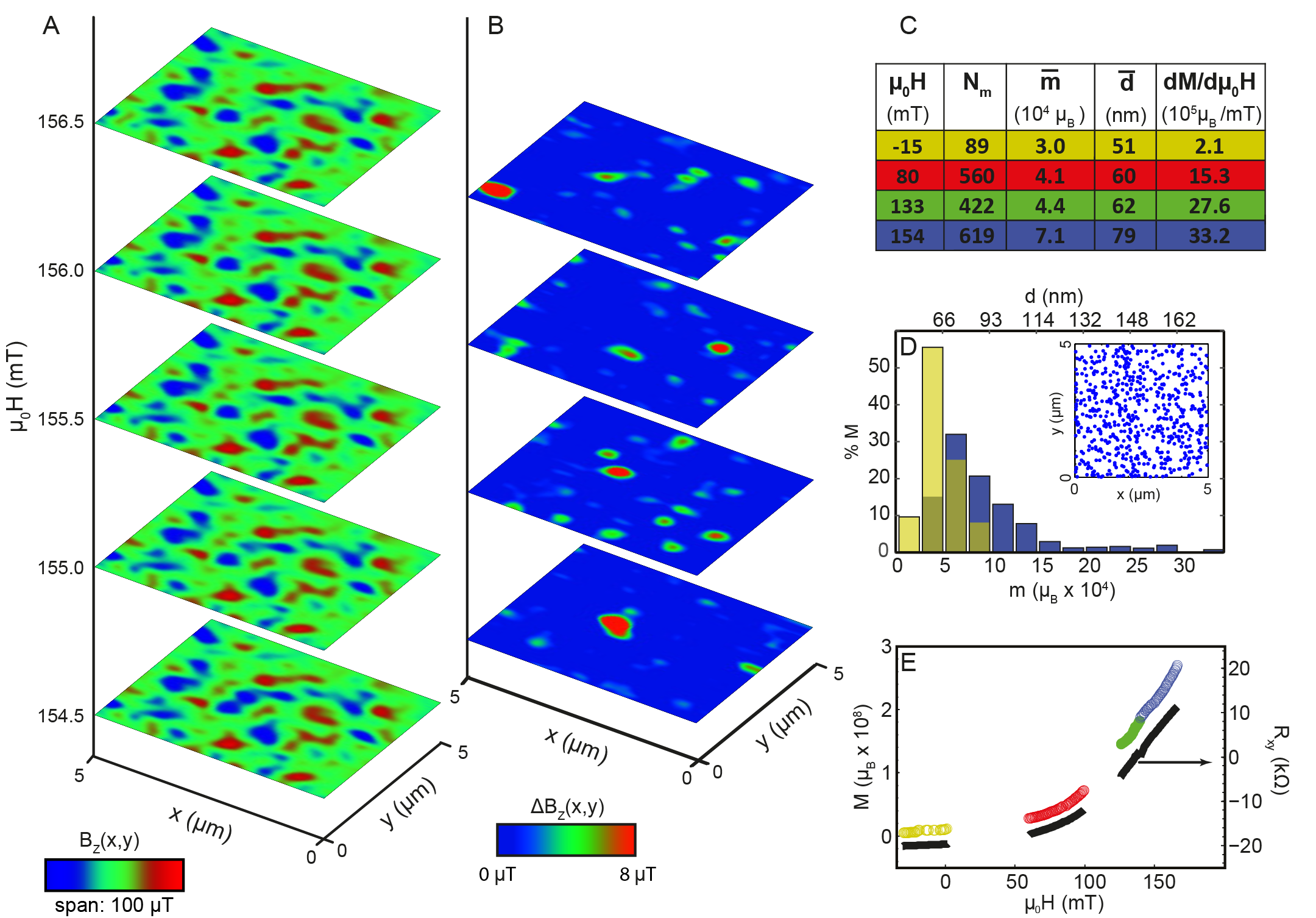}
\caption{\textbf{Magnetization reversal dynamics.}
(A) Sequence of SOT magnetic images $B_z(x,y)$ taken at consecutive magnetic fields in 0.5 mT steps at $T=250$ mK. (B) Differential images $\Delta B_z(x,y)$ obtained by subtracting pairs of consecutive $B_z(x,y)$ images in (A) showing the isolated magnetic reversal events (red) of the superparamagnetic moments (see Movie S1). (C) Statistical analysis of 1690 reversal events attained over ranges of magnetic fields centered around four $\mu_0H$ values: total number of moment reversals $N_{m}$, average magnetic moment $\overline{m}$, average superparamagnetic island diameter $\overline{d}$, and rate of the magnetization change d$M/$d$(\mu_0H)$ over the given range. (D) Chart of relative contribution of different moment sizes $m$ to the total magnetization change $M$ within two field ranges centered at $\mu_0H=-15$ mT (yellow) and $\mu_0H =154$ mT (blue). Inset: location of the moment reversals within the field range around $\mu_0H =154$ mT.
(E) Cumulative magnetization change $M$ due to moment reversals $m$ over four field ranges (left axis, colored symbols) and the simultaneously acquired $R_{xy}$ (right axis, black). The total magnetization in each range is offset by an arbitrary constant.}
	\end{center}
\vspace{0mm}
\end{figure*}
%%%%%%%%%%%%%%%%%%%%%%%%%%%%%%%%%%%%%%%%%%%%%%%
%

%%%%%%%%%%%% FIGURE 3 %%%%%%%%%%%%%%%%%%%%%%%%%%%
\begin{figure*}[ht!]
	\begin{center}
\includegraphics[scale=1]{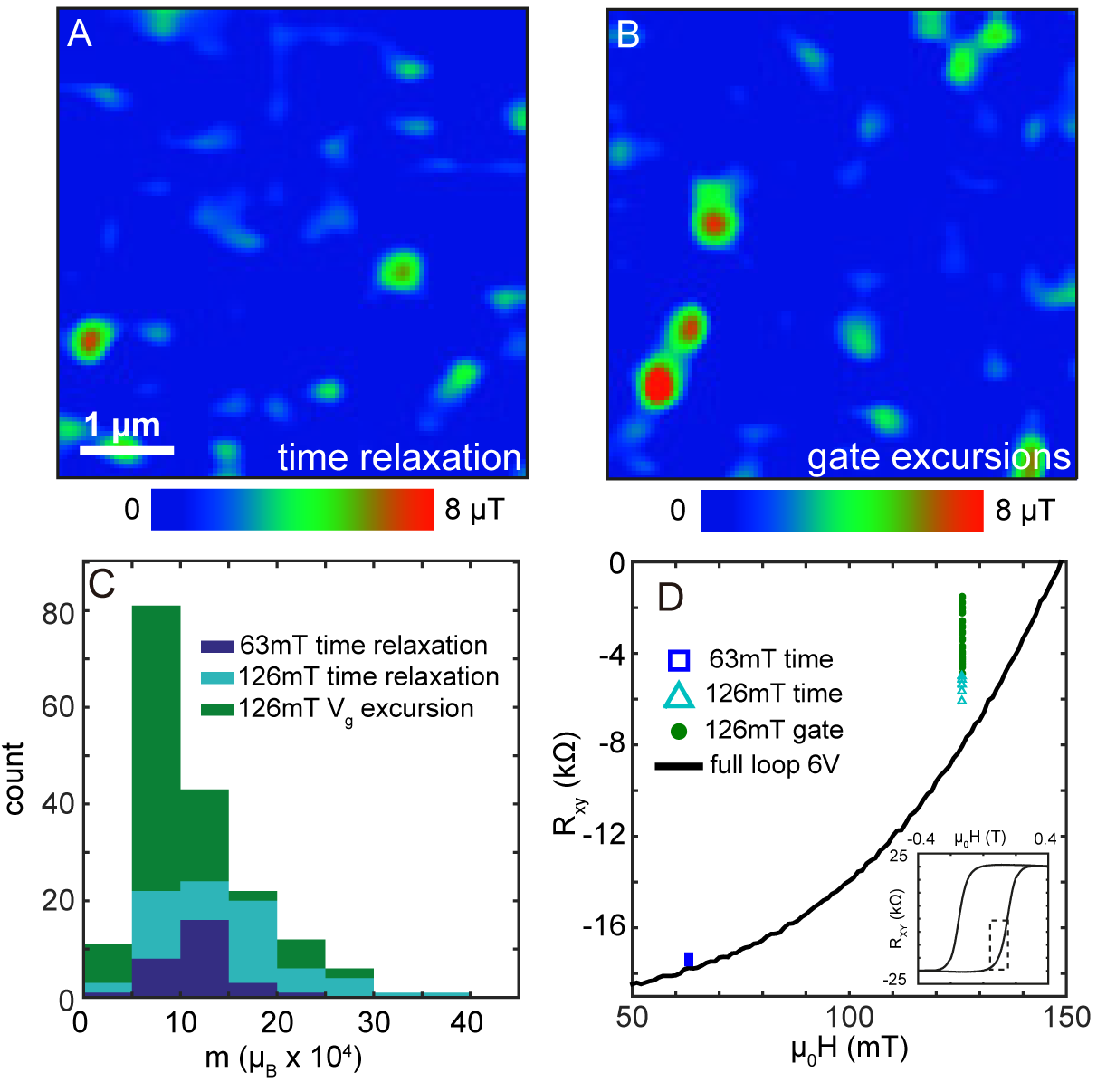}
\caption{\textbf{Temporal and back gate induced relaxation of the superparamagnetic state.}
(A) Differential image $\Delta B_z(x,y)$ obtained by subtraction of two consecutive images acquired at constant $\mu_0H_{set}=126$ mT and $V_g=6$ V after a field ramp from -1 T. Image acquisition time is 200 sec with 50 sec wait time between images. (B) Same as (A) with gate excursion progressively increasing from $\Delta V_g=0.1$ to $1.1$ V in-between consecutive images. (C) Histogram of the temporal relaxation process showing the moment reversals $m$ attained from four consecutive $\Delta B_z(x,y)$ images at $\mu_0H_{set}=63$ mT (dark blue) and at $\mu_0H_{set}=126$ mT (light blue), and of $V_g$-induced relaxation at $\mu_0H_{set}=126$ mT acquired following the temporal relaxation of 20 minutes (green). (D) $R_{xy}$ as a function of field (black) and during relaxation at fixed field taken simultaneously with the magnetic imaging. Temporal relaxation over 20 minutes is more pronounced at 126 mT (light blue) than at 63 mT (dark blue). $V_g$ excursions (green) induce large relaxation at 126 mT. Inset: full $R_{xy}$ hysteresis loop showing the region of interest. }
	\end{center}
\vspace{0mm}
\end{figure*}
%%%%%%%%%%%%%%%%%%%%%%%%%%%%%%%%%%%%%%%%%%%%%%%%%
%

%%%%%%%%%%%% FIGURE 4 %%%%%%%%%%%%%%%%%%%%%%%%%%%
\begin{figure*}[ht!]
	\begin{center}
\includegraphics[width=6.5 in]{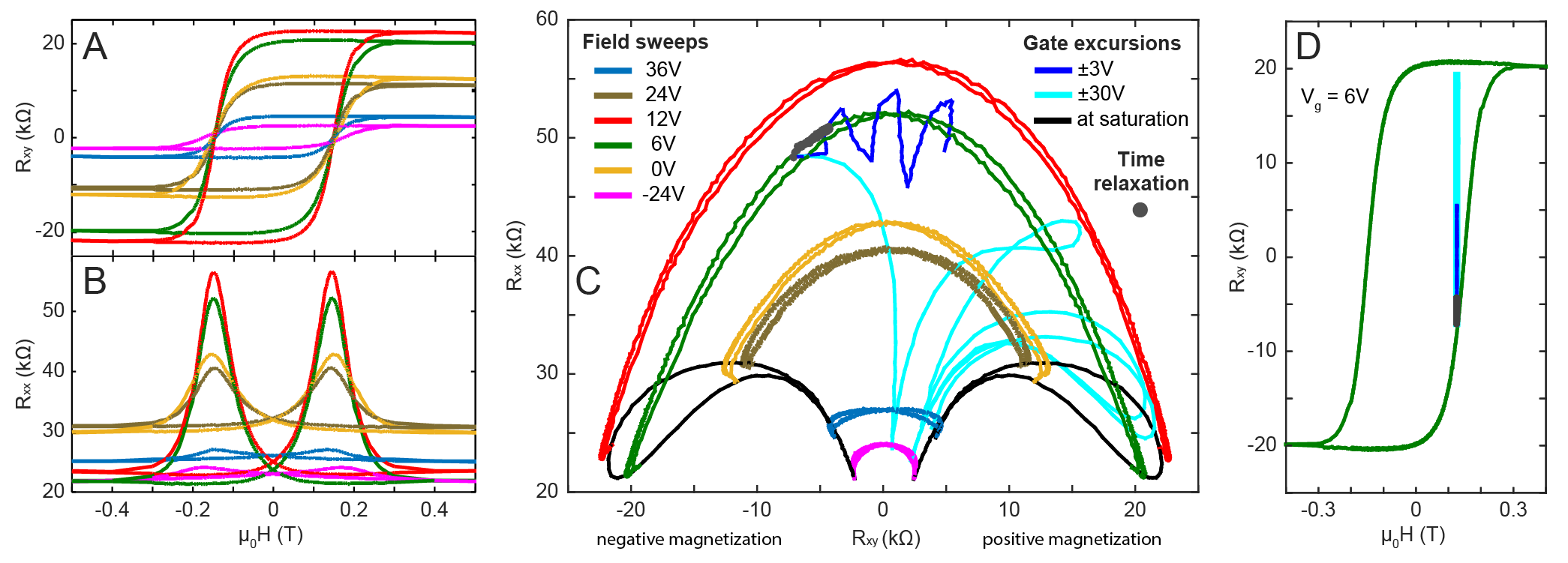}
\caption{\textbf{Transport measurements and universal plot of magnetic relaxation.}
(A) $R_{xy}$ and (B) $R_{xx}$ vs. applied field at $T=250$ mK and different $V_g$ showing magnetic hysteresis with similar $H_c$. (C) The same data plotted as universal arc-like curves of $R_{xx}$ vs. $R_{xy}$ at various $V_g$. Extrema of the arcs correspond to saturation magnetization at -1 T (+1 T) on the lower left (right) end of each arc. Gray dots indicate 60 min of temporal relaxation at $\mu_0H_{set}=126$ mT and $V_g=6$ V (see also Fig. S9). Gate sweeps at $\mu_0H=\pm1$ T (black lines) trace the ends of the arcs, and are reversible. Gate sweep at 126 mT (blue and cyan) are metastable, inducing magnetic relaxation and propagation along the arcs from $R_{xy} < 0$ towards positive saturation. (D) The $R_{xy}$ relaxation data (gray, blue, cyan) and the $R_{xy}$ field sweep at $V_g=6$ V (green).}
	\end{center}
\vspace{0mm}
\end{figure*}
%%%%%%%%%%%%%%%%%%%%%%%%%%%%%%%%%%%%%%%%%%%%%%%%%

\end{document}

\baselineskip24pt
\maketitle

\tableofcontents

\pagebreak
\section{Movies of the magnetic moment reversal dynamics}

\begin{figure}[H]
\includegraphics[width=\textwidth]{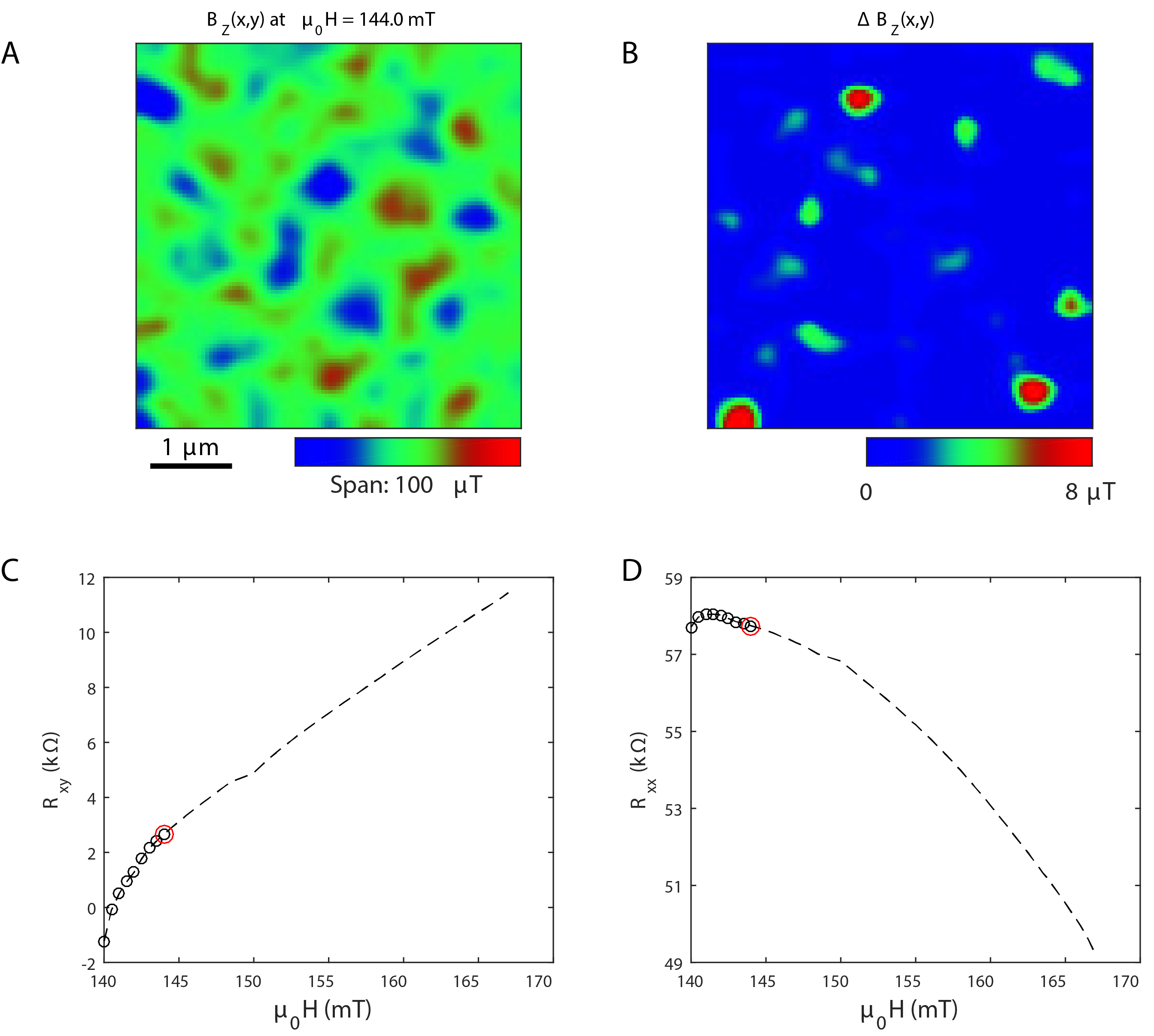}
\caption{\textbf{Movie snapshot of magnetization reversal process in 7QL Cr$_{0.1}$(Bi$_{0.5}$Sb$_{0.5}$)$_{1.9}$Te$_3$ film at T=250 mK.}
(A) Scanning SOT image $5\times5$ $\mu$m$^2$ of the out-of-plane magnetic field $B_z(x,y)$ at $\sim 350$ nm above the sample surface. The running title indicates the value of the applied field which is increased in steps of 0.5 mT between consecutive images (see Movie S1). Image acquisition time was 200 seconds with pixel size of 50 nm. The median background value of the signal is subtracted in each image. (B) Differential image $\Delta B_z(x,y)$ obtained by subtracting the preceding $B_z(x,y)$ image from the current one in (A). Isolated magnetic reversal events of the superparamagnetic moments are visible as red-green peaks. The characteristic size of these peaks is determined by the size of our SOT (200 nm) and by the height of the SOT above the sample, while the physical size of the flipping FM islands is only few tens of nm (see Fig. 2) and is below our resolution. The moments flip in the positive direction of the ascending applied field giving rise to the positive value of the local $\Delta B_z(x,y)$ field peaks. Note the much smaller span of the color code in (B) as compared to (A). (C) $R_{xy}$ of the sample measured simultaneously with the magnetic imaging. The red circle shows the current value while the black circles are the data at previous field values. (D) Same as (C) for $R_{xx}$. Figures 2A and 2B in the main text show few images from the movie while the statistical analysis of the flipping moments in the movie are presented in Figs. 2C-E for field range $\mu_0H=154$ mT.  
}
\label{7QLMovieFrame}
\end{figure}

\begin{figure}[H]
\includegraphics[width=\textwidth]{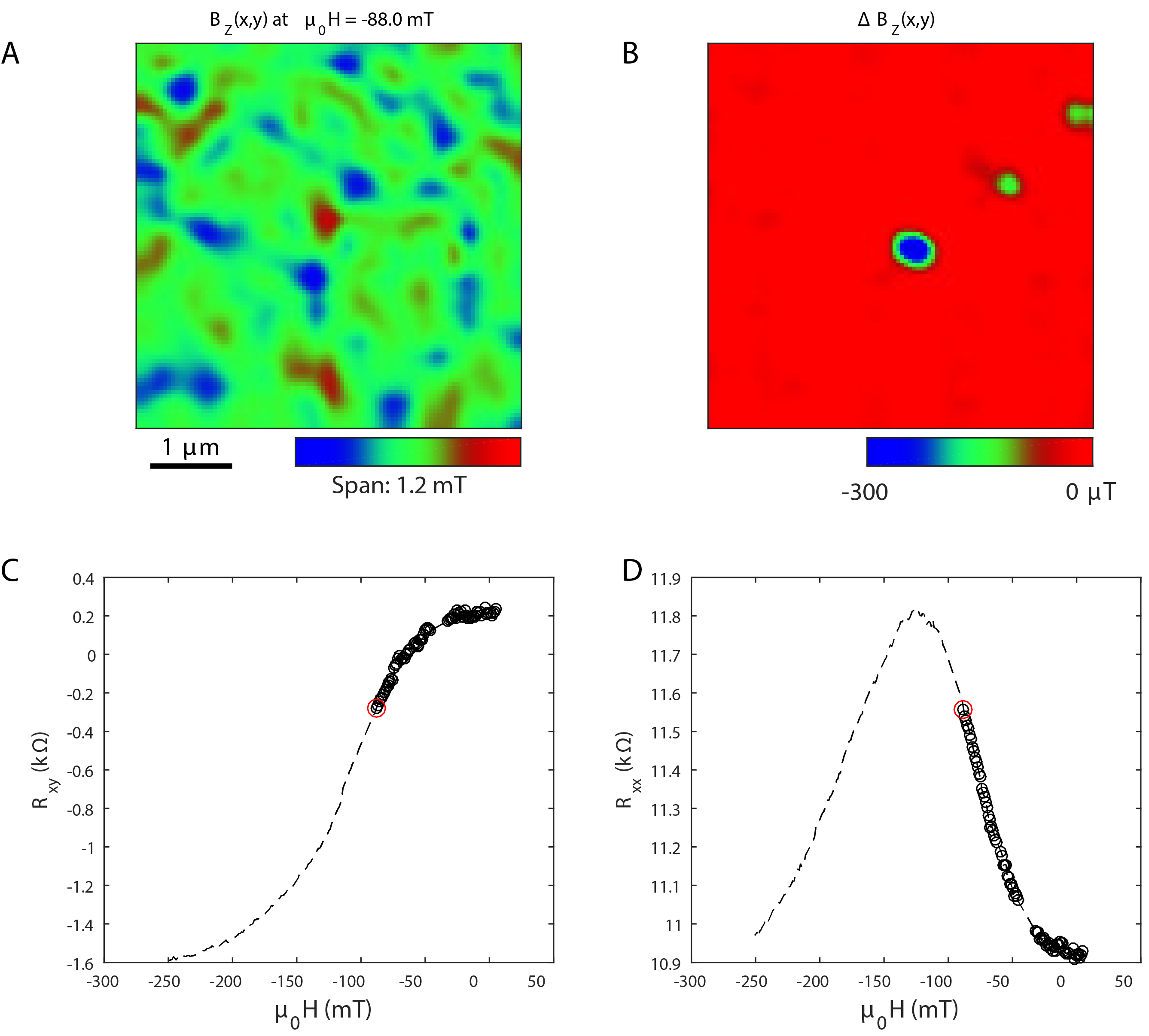}
\caption{\textbf{Movie snapshot of magnetization reversal process in 10QL Cr-doped (Bi,Sb)$_2$Te$_3$ film at T=250 mK.}
(A) Scanning SOT image $5\times5$ $\mu$m$^2$ of the out-of-plane magnetic field $B_z(x,y)$ similar to Fig. S1 but on a different sample and on descending magnetic field in steps of 1 mT (see Movie S2). Image acquisition time was 180 sec with pixel size of 50 nm. (B) Differential image $\Delta B_z(x,y)$ obtained by subtracting a previous $B_z(x,y)$ image from the one in (A). The superparamagnetic moments flip in the direction of the descending magnetic field giving rise to the negative values of the local peaks in $\Delta B_z(x,y)$. (C) $R_{xy}$ of the sample measured simultaneously with the magnetic imaging. (D) Same as (C) for $R_{xx}$.  
}
\label{10QLMovieFrame}
\end{figure}

The fabrication of the SOT devices and the scanning SOT magnetic imaging were performed as described in Refs.%~\cite{Finkler2012, Vasyukov2013}.
 \itshape{ 21,22}\normalfont. 
The scanning magnetic images of $B_z(x,y)$ and the differences between consecutive images $\Delta B_z(x,y)$ were acquired at constant increments of applied magnetic field and compiled into movies (Movies S1 and S2). Figures \ref{7QLMovieFrame} and \ref{10QLMovieFrame} show representative frames from the movies of two Cr-doped (Bi,Sb)$_2$Te$_3$ samples.

%%%%%%%%%%%%%%%%%% FIGURE S1 %%%%%%%%%%%%%%%%%%%%%%%%%%%
%%%%%%%%%%%%%%%%%%%%%%%%%%%%%%%%%%%%%%%%%%%%%%%%%%%%%%
%
%%%%%%%%%%%%%%%%% FIGURE S2 %%%%%%%%%%%%%%%%%%%%%%%%%%%
%%%%%%%%%%%%%%%%%%%%%%%%%%%%%%%%%%%%%%%%%%%%%%%%%%%%%%

\pagebreak
\section{Magnetic moment fitting procedure}

The differential images $\Delta B_z(x,y)$ were fitted by a distribution of point-like magnetic moments with out-of-plane polarization as described in Fig. S3.

%%%%%%%%%%%%%%%%% FIGURE S3 %%%%%%%%%%%%%%%%%%%%%%%%%%%
\begin{figure}[H]
\includegraphics[width=\textwidth]{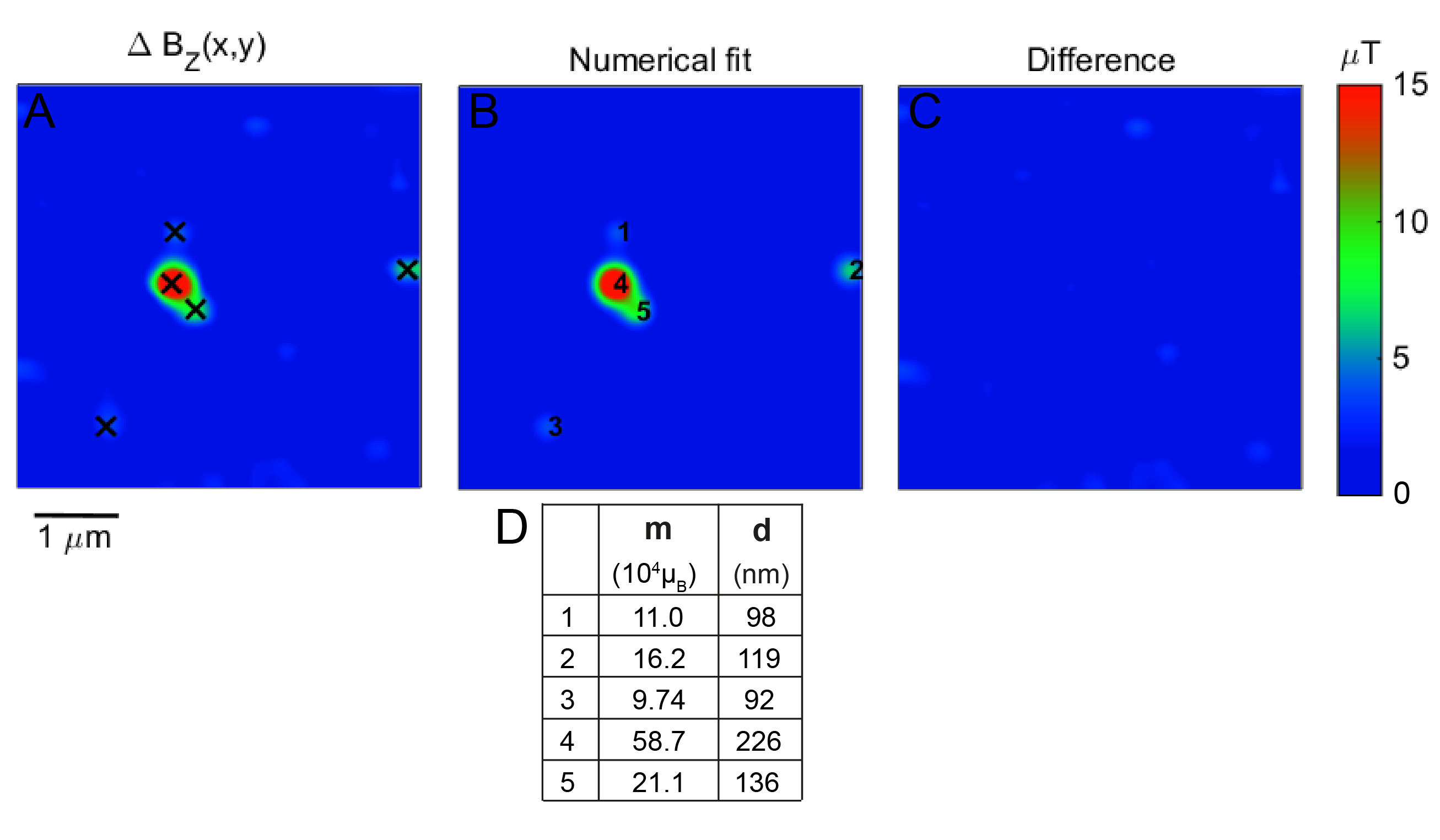}
\caption{\textbf{Fitting procedure of magnetic moments}.
(A) Differential image $\Delta B_z(x,y)$ in 7QL Cr$_{0.1}$(Bi$_{0.5}$Sb$_{0.5}$)$_{1.9}$Te$_3$ film obtained by subtracting $B_z(x,y)$ image at $\mu_0H=154.0$ mT from image at $\mu_0H=154.5$ mT (same as the bottom image in Fig. 2B). Numerical procedure is then used to find local peaks marked by $\times$ that are above a threshold of 4 $\mu$T set by the measurement noise level. (B) Best fit procedure is then applied in which each peak (five in this case) is fitted to the magnetic field generated by a point dipole $2m$ at height $h$ below the SOT convoluted with the active area of the SOT (the factor 2 arises from the fact that the superparamagnetic island magnetization flips from $-m$ to $+m$). (C) The quality of the fit is demonstrated by subtracting the fit (B) from the image (A). (D) The resulting magnitudes $m$ of the fitted dipoles at a fixed optimal $h=345$ nm. The diameter $d$ of the superparamagnetic islands is then calculated from their moment $m$ assuming an average saturation magnetization of $3 \mu_\textrm{B} / \textrm{Cr}$ atom (obtained from global magnetization measurements presented in Fig.~\ref{giantSQUID}), Cr doping of 5\%, unit cell size of 0.43 nm, and thickness of 7 nm. The resulting size of the islands is smaller than SOT diameter or the height $h$ justifying the point dipole aproximation. A full simulation taking into account the finite size of the islands gave similar results.
}
\label{dipoleFitOriginal}
\end{figure}
%%%%%%%%%%%%%%%%%%%%%%%%%%%%%%%%%%%%%%%%%%%%%%%%%%%%%%

\pagebreak
\section{Spatial distribution of the magnetization reversals}

%%%%%%%%%%%%%%%%% FIGURE S4 %%%%%%%%%%%%%%%%%%%%%%%%%%%
\begin{figure}[H]
\includegraphics[width=\textwidth]{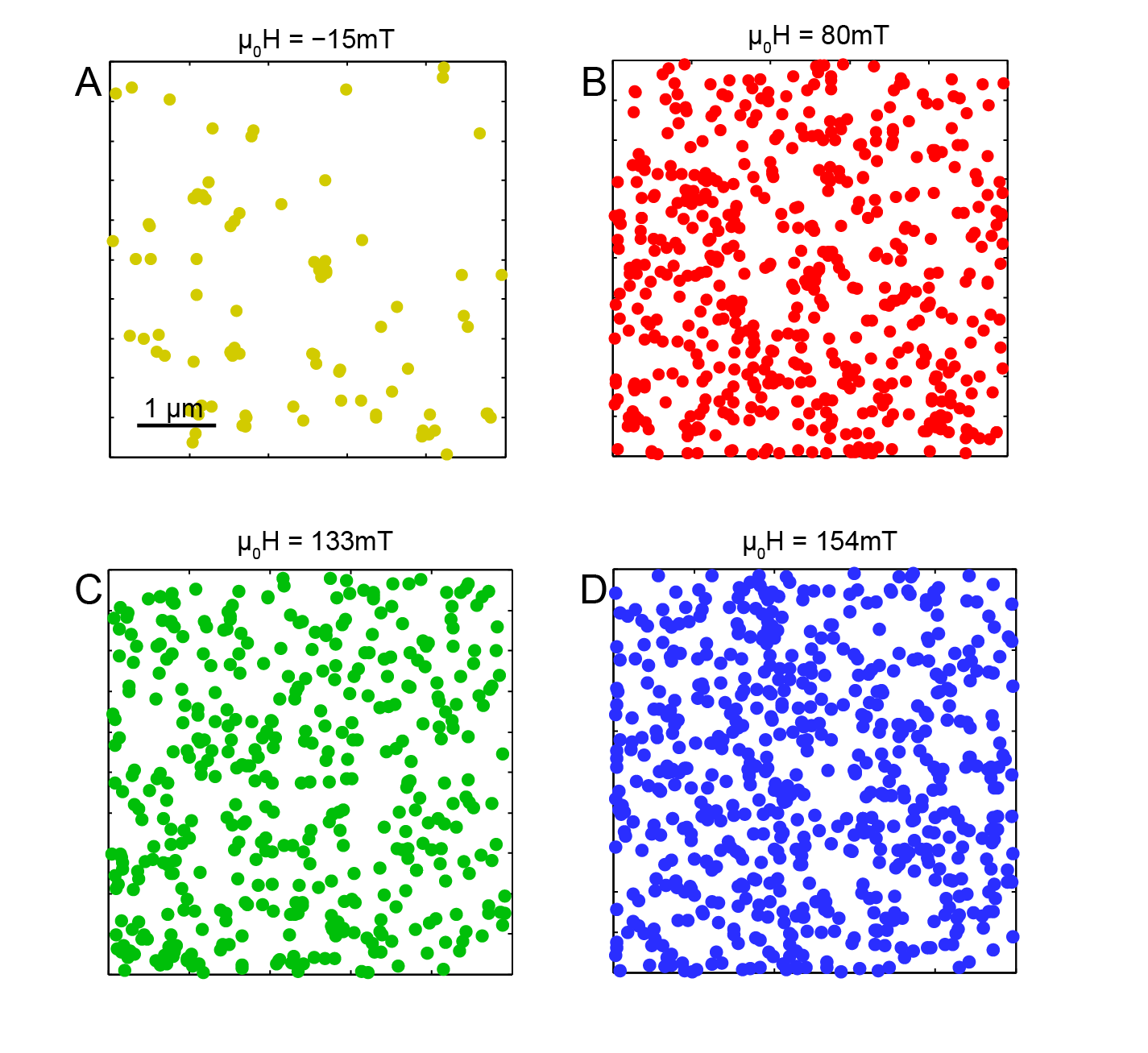}
\caption{\textbf{Spatial distribution of the magnetization reversal process in 7QL Cr$_{0.1}$(Bi$_{0.5}$Sb$_{0.5}$)$_{1.9}$Te$_3$ sample at T=250 mK.} 
(A-D) Location of all the moments that flipped upon increasing the applied field from $\mu_0H=-30$ mT to 1 mT (A), $\mu_0H=60$ to 100 mT (B), $\mu_0H=126$ to 140 mT (C), and $\mu_0H=140$ to 167 mT (D) corresponding to the four field ranges in Figs. 2C-E. The random distribution of the moment locations further demonstrates that the magnetization transition occurs through reversal of weakly interacting superparamagnetic islands rather than through FM domain wall motion.
}
\label{dipoleDist}
\end{figure}
%%%%%%%%%%%%%%%%%%%%%%%%%%%%%%%%%%%%%%%%%%%%%%%%%%%%%%

\pagebreak
\section{$V_g$ dependence of the dynamics of magnetic relaxation}

Figures 4A,B show that $V_g$ has only a small effect on the coercive field $H_c$ while $\Delta V_g$ excursions near $H_c$ relax the magnetization dramatically as shown in Figs. 4C,D. We have also investigated the temporal relaxation process at different constant values of $V_g$ as shown in Fig. S5.

%%%%%%%%%%%%%%%% FIGURE S5 %%%%%%%%%%%%%%%%%%%%%%%%%%%
\begin{figure}[H]
\includegraphics[scale = 1]{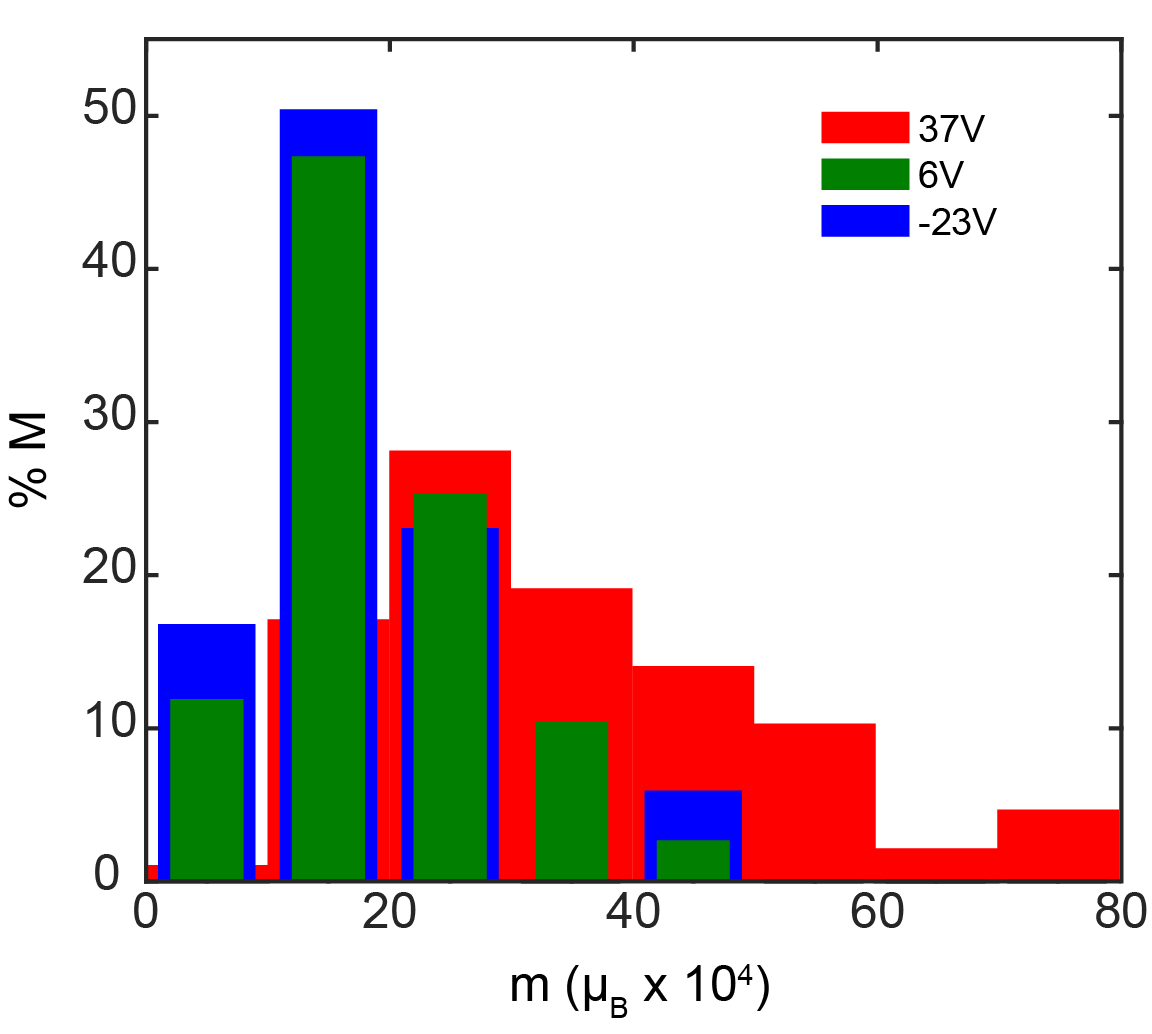}
\caption{\textbf{Statistical analysis of temporal moment relaxation for different $V_g$ values in 7QL Cr$_{0.1}$(Bi$_{0.5}$Sb$_{0.5}$)$_{1.9}$Te$_3$ film at T=250 mK.} 
The histogram shows the relative contribution of moments of different magnitudes $m$ to the total magnetization relaxation $M$ for three values of $V_g$. For each histogram the magnetic state was initialized at a fixed $V_g$ by sweeping the applied field from -1 T to 125 mT and consecutive $B_z(x,y)$ images were acquired for 25 min at fixed $\mu_0H_{set}=125$ mT similar to Fig. 3A. While for $V_g=-23$ and 6 V the distributions are similar, at $V_g=37$ V a clear shift of the distribution to larger moments is visible showing that at the same applied field the dynamics of the relaxation process has some dependence on $V_g$. These results indicate that the magnetic anisotropy energy of individual islands is affected by $V_g$ apparently through variations in the disorder-induced random local potential or density of states. Since the local variations are random their average effect on the global coercive field could be small as observed in Figs. 4A,B.
}
\label{diffGates}
\end{figure}
%%%%%%%%%%%%%%%%%%%%%%%%%%%%%%%%%%%%%%%%%%%%%%%%%%%%%

\pagebreak
\section{Transport measurements}

The samples were mechanically patterned into a Hall-bar geometry and the transport measurements were performed using three synchronized SR830 lock-in amplifiers for simultaneous measurement of the current, longitudinal voltage, and transverse voltage (Fig. S6). The applied current was 10 to 20 nA at a frequency of 17.71 Hz. In order to minimize the mixing between longitudinal and transverse resistances switching between the current and voltage leads in VdP configuration was performed using Keithly7001 switch box for most of the presented data. The gate voltage was applied using a DC source (Keithley2400). The sweeping rate of $V_g$ was limited to ~0.5 V$/$sec to ensure absence of heating, and after reaching the target $V_g$ there was a 3 seconds waiting time before the measurement.  

The simultaneous magnetotransport and SOT imaging measurements were performed in Oxford Heliox $^3$He cryostat equipped with 12 T magnet. In order to prevent heating the magnetic field ramping was kept at a low constant rate of 0.4 T/min in all experiments. This rate was used for magnetotransport measurements as well as for preparation of the initial magnetization state for temporal and gate-induced magnetic relaxation measurements. Figures S7 to S9 show the dependence of transport coefficients on temperature, back gate voltage, and the temporal relaxation. Since large $\Delta V_g$ excursions cause pronounced relaxation of the magnetization we have also compared the hysteretic behavior of the transport coefficients in presence of gate excursions as shown in Fig.~\ref{gateShaking}.

%%%%%%%%%%%%%%%%% FIGURE S6 (new) %%%%%%%%%%%%%%%%%%%%%
\begin{figure}[H]
\includegraphics[width=\textwidth]{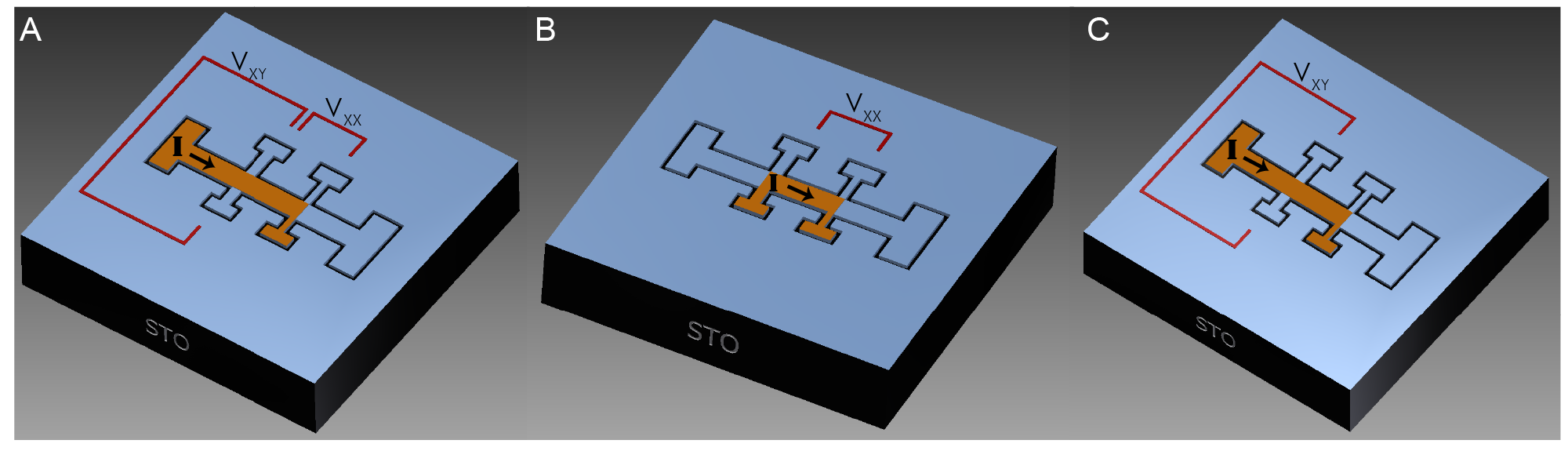}
\caption{\textbf{Schematics of transport measurements.} 
Three panels showing the scheme of the scratched Hall bar appearing in Fig. 1C. The width of the central channel is 500 $\mu m$, the inner distance between two voltage probes on the same side is 1000 $\mu m$ and the width of each voltage channel is 200 $\mu m$. The orange part marks the two current leads and the channel between them i.e. where bulk current flows. The red lines indicate measurement configuration of $V_{xx}$ or $V_{xy}$. 
(A) Hall bar measurement configuration for simultaneous measurements of $I$, $V_{xx}$ and $V_{xy}$ using three lock-in amplifiers resulting in $R_{xx} = \frac{V_{xx}}{I}$ and $R_{xy} = \frac{V_{xy}}{I}$.
(B) VdP configuration for measurement of $R_{xx}$ with switching between current and voltage leads using a switch box.
(C) VdP configuration with switching for measuring $R_{xy}$.
}
\label{measurementScheme}
\end{figure}

%%%%%%%%%%%%%%%%% FIGURE S7 %%%%%%%%%%%%%%%%%%%%%%%%%%%
\begin{figure}[H]
\includegraphics[width=\textwidth]{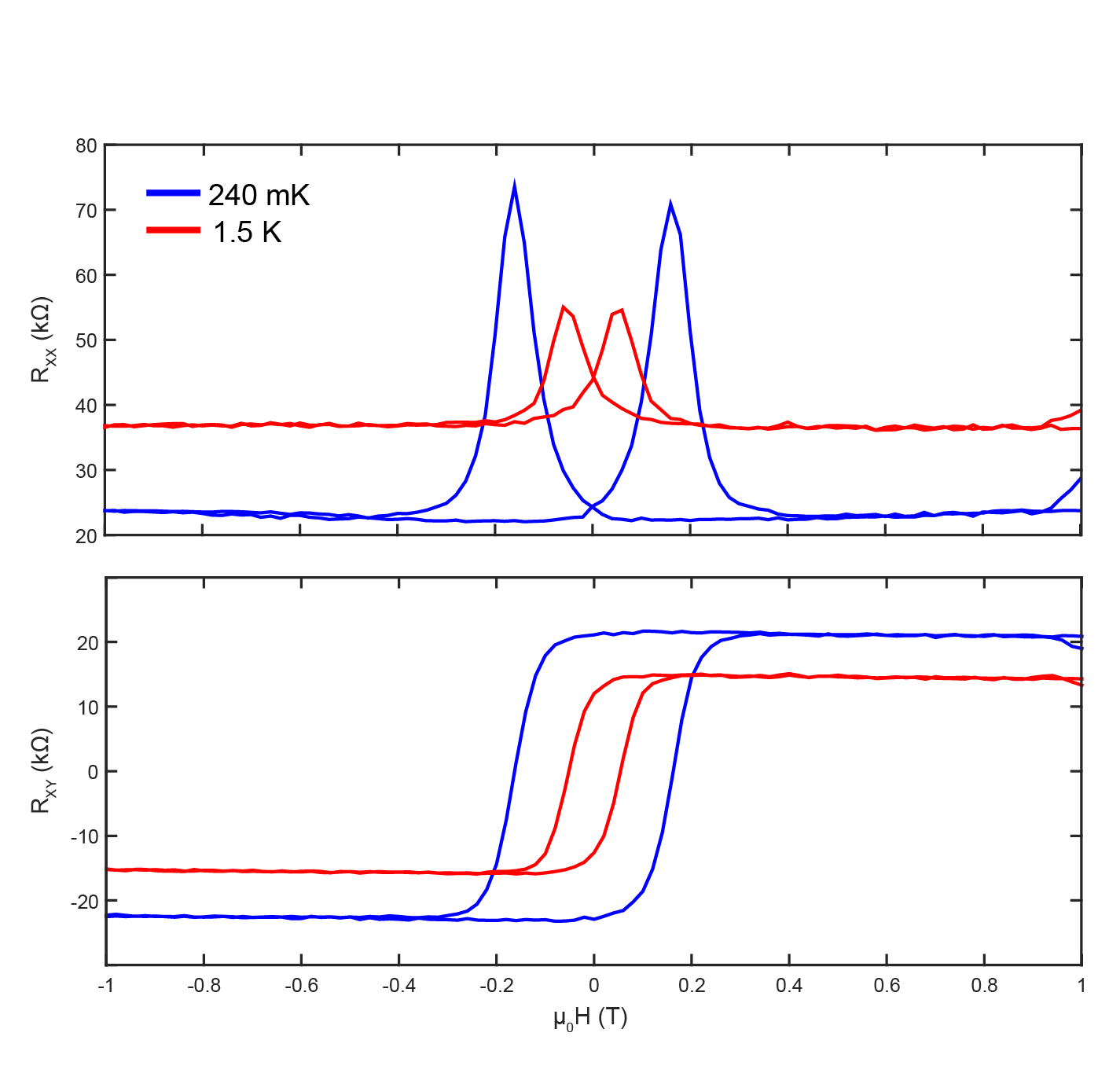}
\caption{\textbf{Temperature dependence of the transport coefficients in 7QL Cr$_{0.1}$(Bi$_{0.5}$Sb$_{0.5}$)$_{1.9}$Te$_3$ film.} 
Hysteretic loops of $R_{xx}$ and $R_{xy}$ vs. applied field at $V_g=8$ V at two temperatures $T=240$ mK and 1.5 K. The coercive field at 1.5 K is reduced to about a third of its value at the base temperature.
}
\label{RvsHvsT}
\end{figure}
%%%%%%%%%%%%%%%%%%%%%%%%%%%%%%%%%%%%%%%%%%%%%%%%%%%%%%

%%%%%%%%%%%%%%%% FIGURE S8 %%%%%%%%%%%%%%%%%%%%%%%%%%%
\begin{figure}[H]
\includegraphics[width=\textwidth]{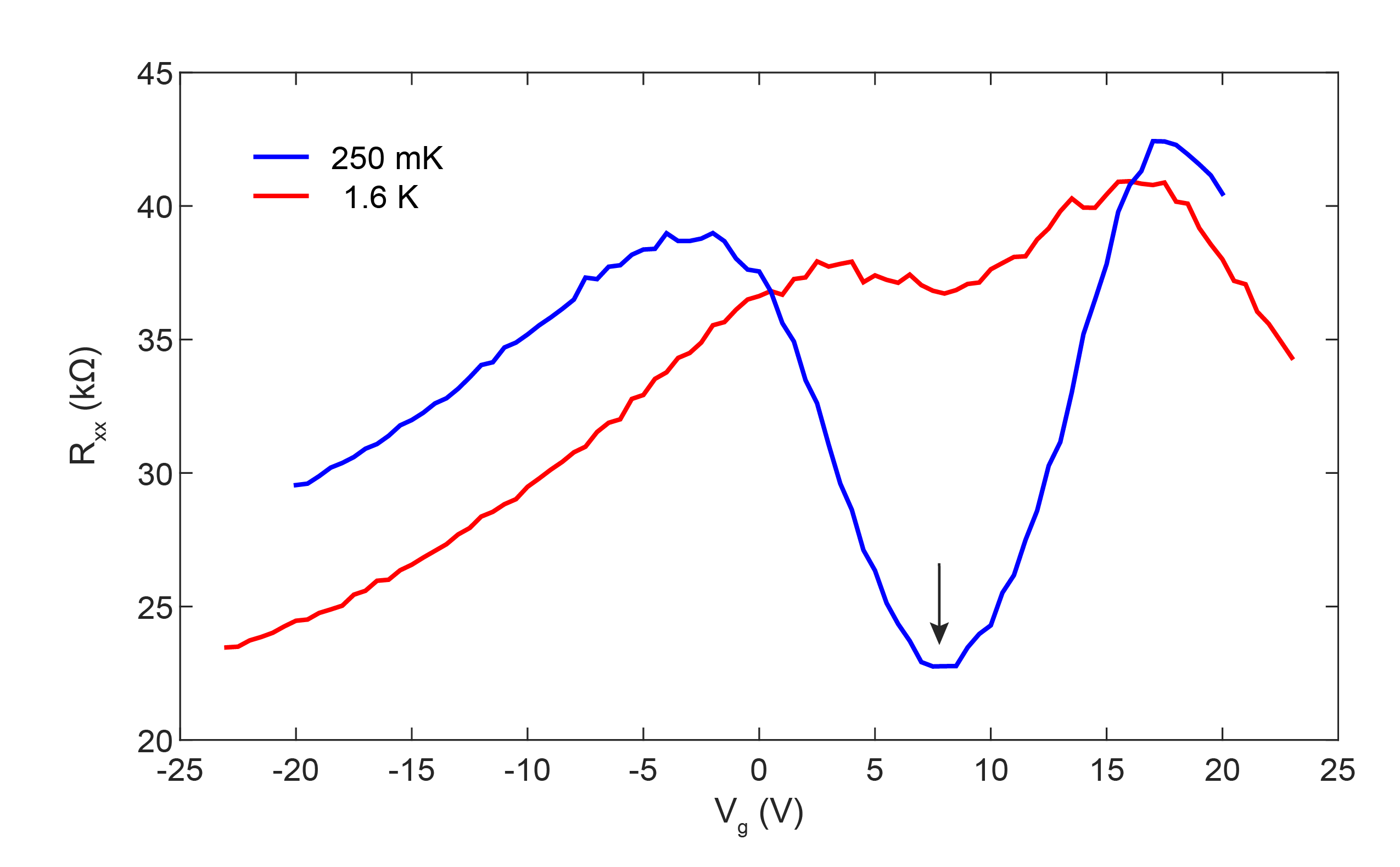}
\caption{\textbf{Gate voltage dependence of $R_{xx}$ in 7QL Cr$_{0.1}$(Bi$_{0.5}$Sb$_{0.5}$)$_{1.9}$Te$_3$ sample.} 
The longitudinal resistance $R_{xx}$ at $\mu_0H=1$ T vs. $V_g$ at $T=250$ mK and 1.6 K. The incipient QAH state is clearly visible at $T=250$ mK as a pronounced dip around $V_g=8$ V marked by the arrow. A smaller dip is resolved also at $T=1.6$ K.
}
\label{RvsGatevsT}
\end{figure}
%%%%%%%%%%%%%%%%%%%%%%%%%%%%%%%%%%%%%%%%%%%%%%%%%%%%%

%%%%%%%%%%%%%%%% FIGURE S9 %%%%%%%%%%%%%%%%%%%%%%%%%%%
\begin{figure}[H]
\includegraphics[width=\textwidth]{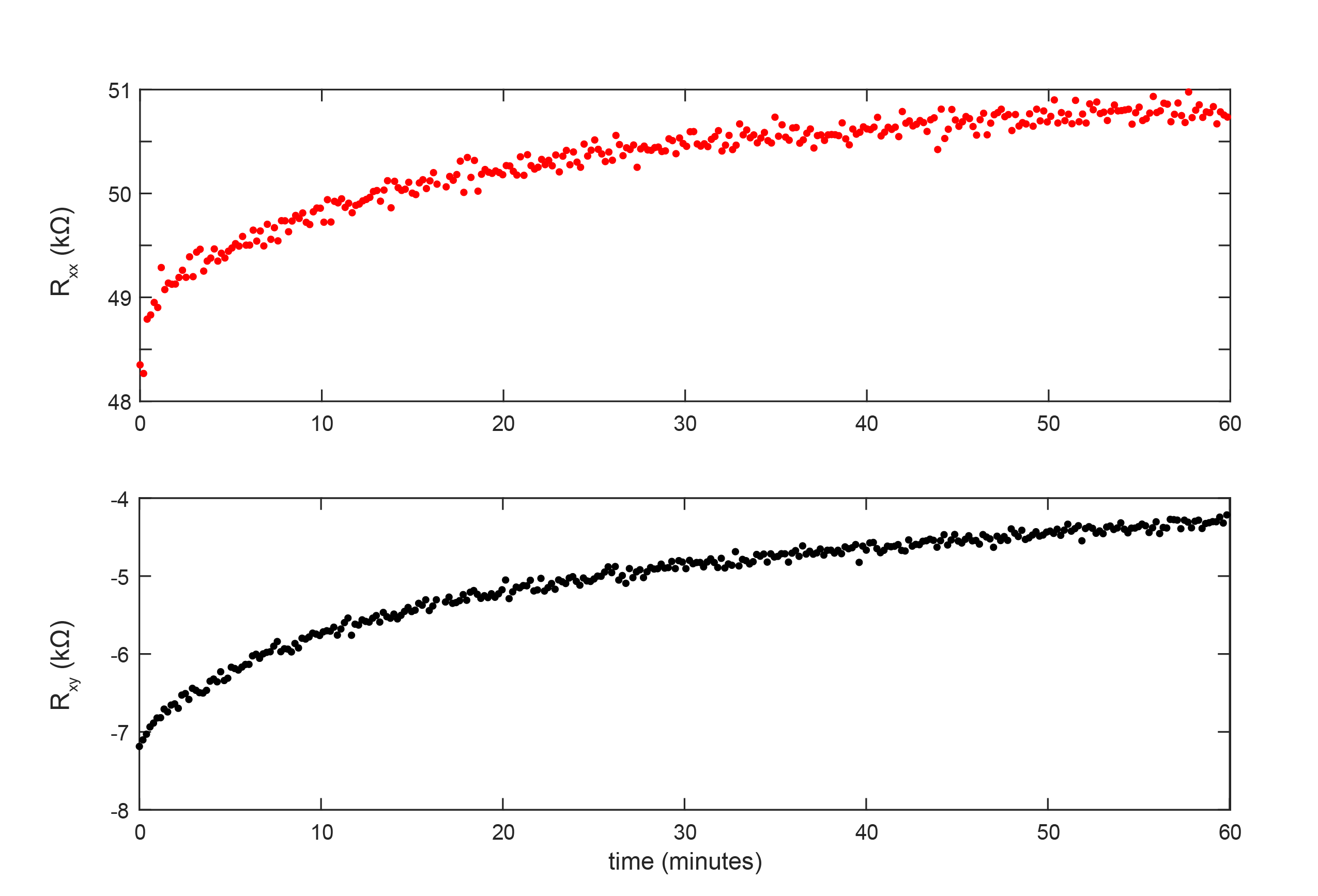}
\caption{\textbf{Temporal relaxation of transport coefficients near the coercive field in 7QL Cr$_{0.1}$(Bi$_{0.5}$Sb$_{0.5}$)$_{1.9}$Te$_3$ sample.} 
Time dependence of $R_{xx}$ and $R_{xy}$ at $\mu_0H_{set}=126$ mT at $V_g=6$ V after ramping the field from -1 T. The relation between $R_{xx}$ and $R_{xy}$ follows the universal arc-like scaling as shown in Fig. 4C by the gray dots.
}
\label{timeRelax}
\end{figure}
%%%%%%%%%%%%%%%%%%%%%%%%%%%%%%%%%%%%%%%%%%%%%%%%%%%%%

%%%%%%%%%%%%%%% FIGURE S10 %%%%%%%%%%%%%%%%%%%%%%%%%%%
\begin{figure}[H]
\includegraphics[width=\textwidth]{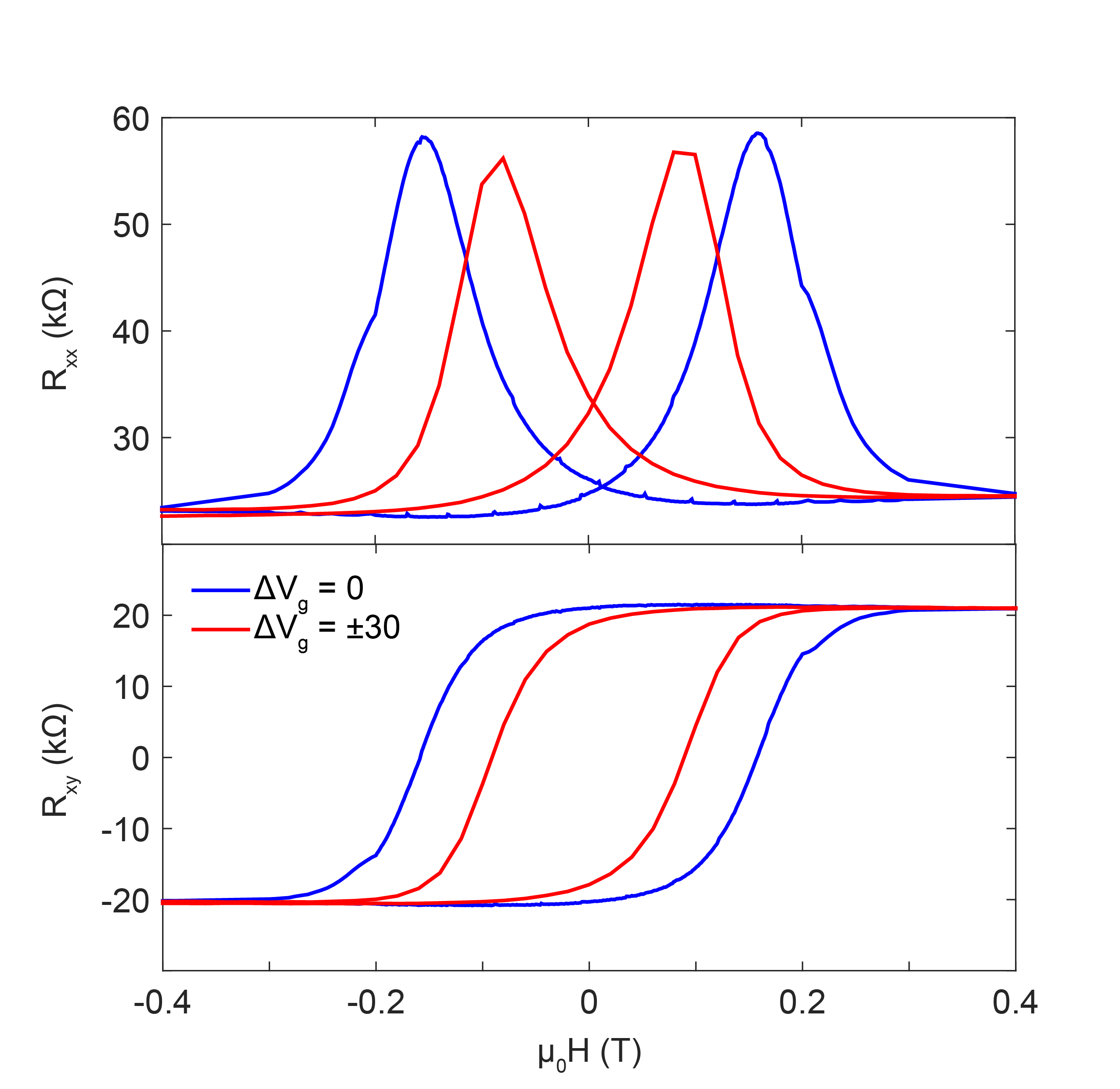}
\caption{\textbf{Transport coefficients in 7QL Cr$_{0.1}$(Bi$_{0.5}$Sb$_{0.5}$)$_{1.9}$Te$_3$ sample with continuous gate excursions.} 
Hysteretic loops of $R_{xx}$ and $R_{xy}$ vs. applied field at $V_g=6$ V without (blue) and with (red) gate excursions. In the former case the transport coefficients were measured continuously while sweeping the applied field. In the latter case at each value of the applied field, gate excursion of $\Delta V_g=\pm 30$ V was applied for ten times prior to the measurement of $R_{xx}$ and $R_{xy}$. Increasing the number of $\Delta V_g$ repetitions did not induce additional change in the transport coefficients. The effective coercive field is seen to be reduced substantially by the gate excursions in contrast to Fig. 4A,B which shows no appreciable change in $H_c$ for field sweeps at various constant gate voltages over the same range of $V_g$ values. Note that the resulting reduction in $H_c$ is much larger than what can be achieved by temporal relaxation (see Figs. ~\ref{timeRelax} and  4D). Despite the large reduction in the hysteresis by the gate excursions a fully reversible equilibrium state could not be reached. 
}
\label{gateShaking}
\end{figure}
%%%%%%%%%%%%%%%%%%%%%%%%%%%%%%%%%%%%%%%%%%%%%%%%%%%%

\pagebreak
\section{Growth and structural characterization}
Samples were grown by molecular beam epitaxy (MBE) on commercial SrTiO$_3$ substrates with dimensions of 5 x 5 x 0.5 mm$^3$. Substrates were annealed in oxygen at ~925$^\circ$ C for 2 hour 30 min typically and were screened using atomic force microscopy (AFM) to make sure the surface was atomically ordered with an RMS roughness of less than 0.3 nm. Substrates were indium mounted for growth in the MBE chamber and outgassed at a substrate temperature of 550$^\circ$ C for about 1 hour to remove any residual contamination before growth. The TI films were grown using elemental materials of at least 5N purity evaporated from Knudsen type thermal cells at a growth rate of about 0.4 QL per minute at a temperature of ~250$^\circ$ C, as measured by a pyrometer. 

Transmission electron microscopy (TEM) studies were carried out on a sample grown under similar conditions. The TEM sample was prepared by focused ion beam (FIB) and measured in an FEI Titan double aberation corrected TEM at 200 keV. High angle annular dark field scanning TEM (HAADF) images showed a well ordered crystal with the expected quintuple layer structure. Two slightly bright lines in the middle of each QL correspond to the location of the heavier Bi atoms. A disordered amorphous region about 1 nm thick was observed at the interface with the SrTiO$_3$. We have observed a similar layer in undoped films as well. Energy dispersive spectroscopy (EDS) was used to map the spatial distribution of elements. Within each of six regions studied the distributions of elements appeared uniform with the exception that Bi appeared deficient in the interfacial region.

%%%%%%%%%%%%%%% FIGURE S11 %%%%%%%%%%%%%%%%%%%%%%%%%%%
\begin{figure}[H]
\includegraphics[width=\textwidth]{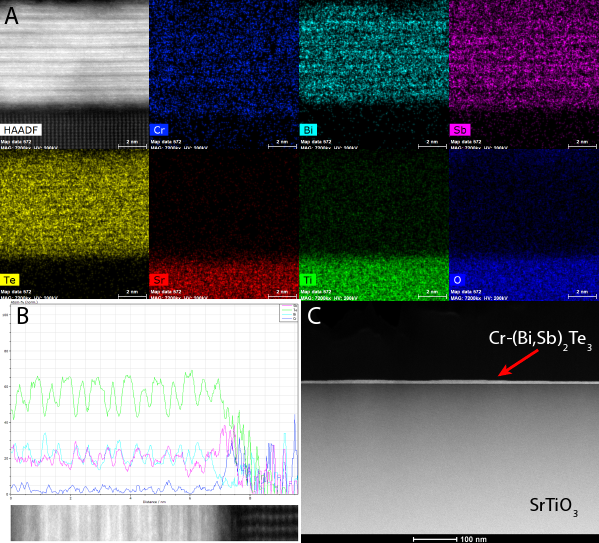}
\caption{\textbf{STEM imaging and EDS elemental mapping of a Cr doped (Bi,Sb)$_2$Te$_3$ film on SrTiO$_3$.}
(A) HAADF image showing the crystal structure and EDS elemental maps of the various atomic species. Less Bi is observed in the disordered layer at the interface with the SrTiO$_3$. (B) Linecut made by averaging over the area shown. The Te-Bi(Sb)-Te-Bi(Sb)-Te QL structure is clearly seen. Te, Bi, Sb and Cr are shown in green, cyan, magenta, and blue respectively. (C) Large scale HAADF image of the film.   
}
\label{TEM}
\end{figure}
%%%%%%%%%%%%%%%%%%%%%%%%%%%%%%%%%%%%%%%%%%%%%%%%%%%%

\pagebreak
\section{Transport and superparamagnetic dynamics in 10QL Cr-doped $(Bi,Sb)_2Te_3$ sample }
\begin{figure}[H]
\includegraphics[scale = 0.8]{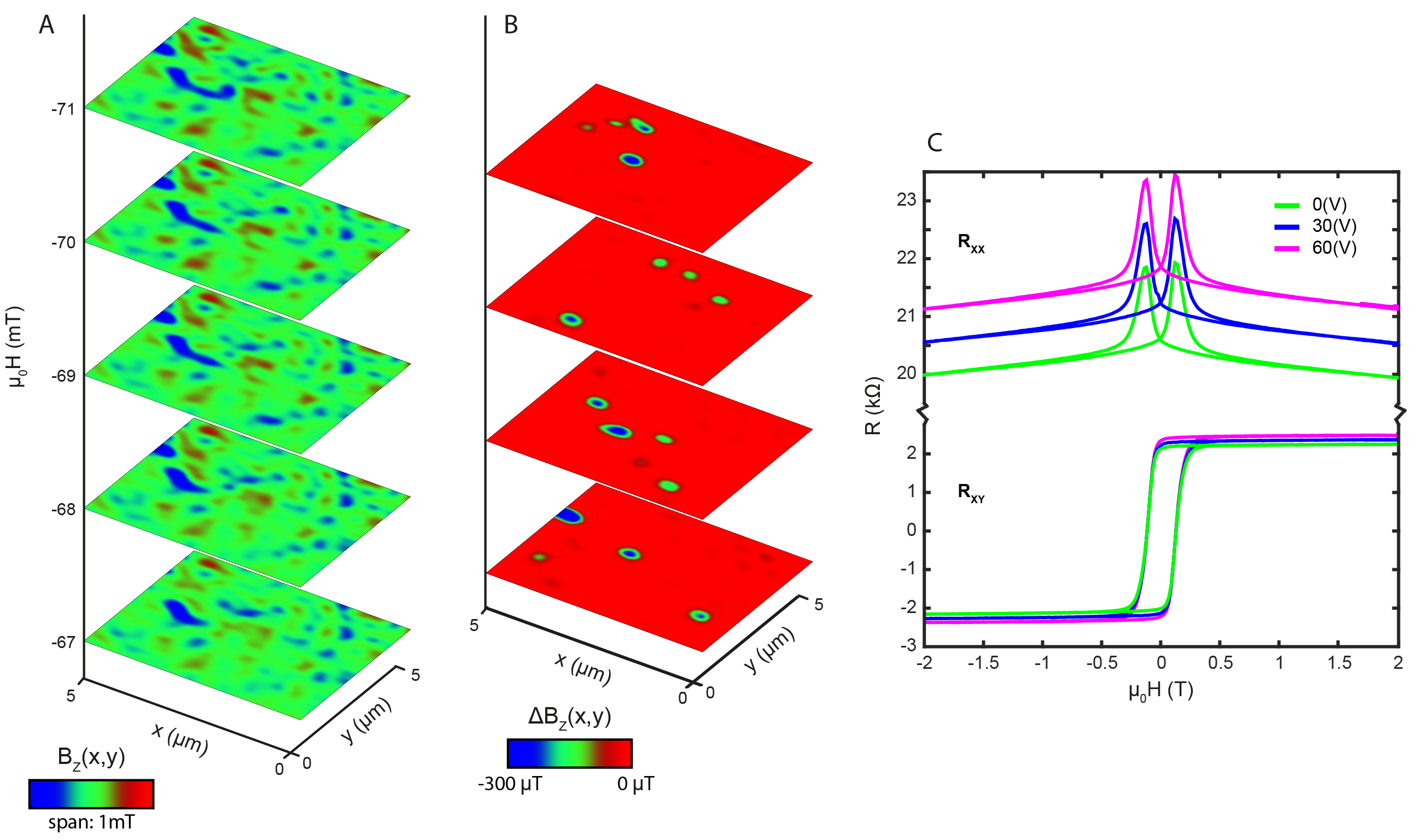}
\caption{\textbf{Magnetization reversal dynamics and transport coefficients in 10QL thick Cr-doped (Bi,Sb)$_2$Te$_3$ thin film at T=250 mK.}
(A) Sequence of $5\times5$ $\mu$m$^2$ scanning SOT magnetic images $B_z(x,y)$ taken at consecutive descending magnetic fields in 1 mT steps. (B) Differential images $\Delta B_z(x,y)$ obtained by subtracting pairs of consecutive $B_z(x,y)$ images in (A) showing the isolated negative magnetic reversal events of the superparamagnetic moments (blue). A larger sequence of images is available in Movie S2 as described in Fig. S2. (C)  Transport coefficients vs. applied field at three values of $V_g$.
}
\label{OtherCrSampleStack}
\end{figure}

\begin{figure}[H]
\includegraphics[width=\textwidth]{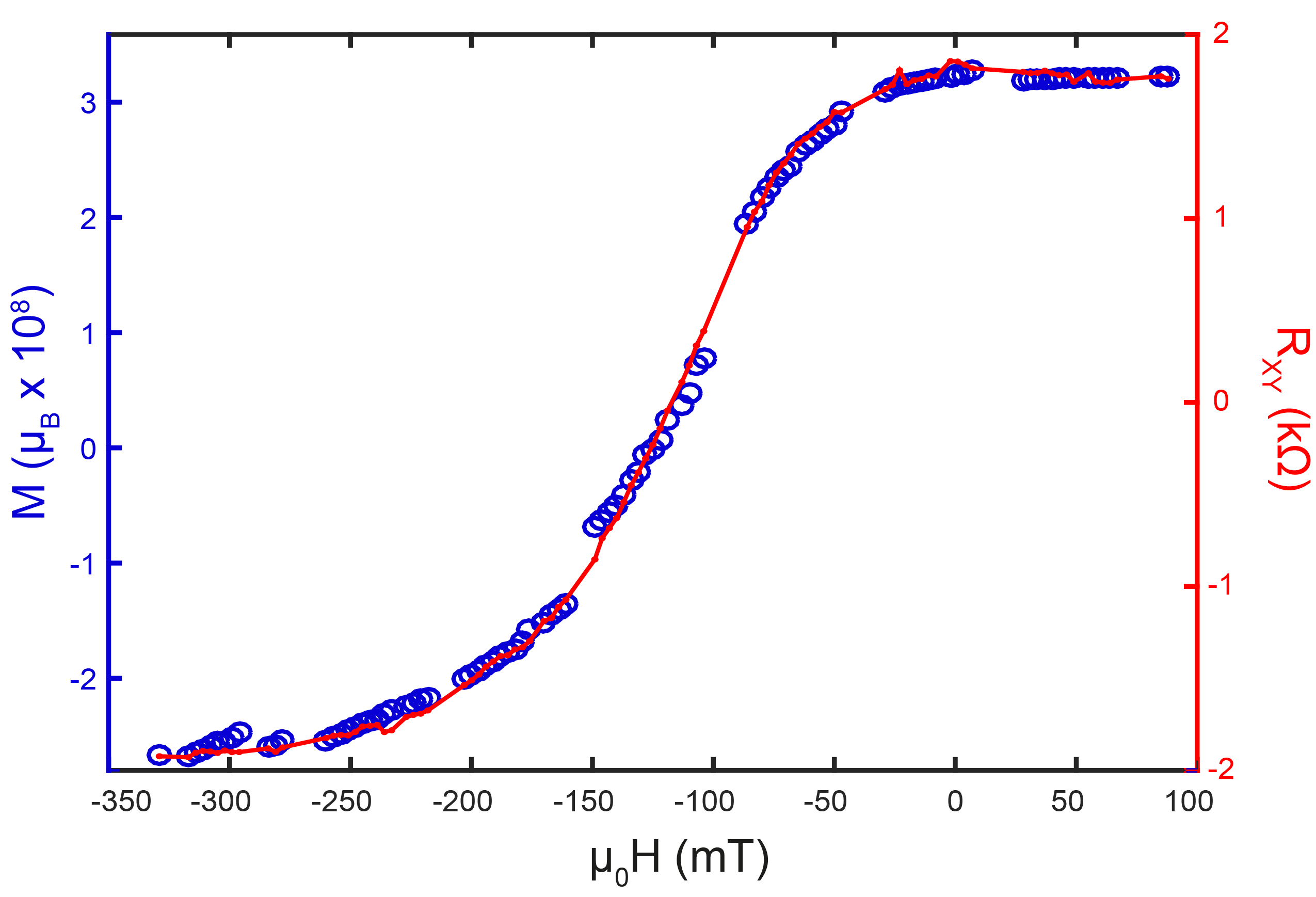}
\caption{\textbf{ Scaling of the cumulative magnetization change and the transverse resistance in 10QL Cr-doped (Bi,Sb)$_2$Te$_3$ film at T=250 mK.}
Evolution of magnetization $M$ (left axis, blue circles) on descending magnetic field (after sweep down from 1 T) obtained by cumulative addition of the superparamagnetic moment reversals $M = \sum{2m}$, derived from a sequence of images like in Fig.~\ref{OtherCrSampleStack} and the simultaneously measured $R_{xy}$ (right axis, red). The data points were acquired only at ranges of fields where the SOT had high sensitivity resulting in the gaps between the ranges. The magnetization in each range is offset by an arbitrary constant. Through most of the transition region a close relation between $R_{xy}$ and $M$ is observed.
}
\label{Rxyvsm}
\end{figure}

Similar studies were performed on an additional 10QL Cr-doped sample. Figures S2, ~\ref{OtherCrSampleStack}, and  ~\ref{Rxyvsm} show that the transport coefficients in this sample are far from QAH quantization, however the superparamagnetic behavior is qualitatively the same.
%%%%%%%%%%%%%%%% FIGURE S12 %%%%%%%%%%%%%%%%%%%%%%%%%%%
%%%%%%%%%%%%%%%%%%%%%%%%%%%%%%%%%%%%%%%%%%%%%%%%%%%%%
%%%%%%%%%%%%%%% FIGURE S13 %%%%%%%%%%%%%%%%%%%%%%%%%%%
%%%%%%%%%%%%%%%%%%%%%%%%%%%%%%%%%%%%%%%%%%%%%%%%%%%%

\pagebreak
\section{Global magnetization studies}
Magnetization measurements were taken of a 40 nm thick sample using a commercial SQUID magnetometer with the applied field perpendicular to the film (Fig. S14). For the temperature sweeps the sample was cooled from 305K to 3.5K in either a 9000 Oe field for the field cooled (FC) curve or 1.3 Oe field for the zero field cooled (ZFC) curve (1.3 Oe is the remenance in the magnet after demagnetization, so this corresponds to zero field). The applied field is then set to 50 Oe and the magnetization is measured as the sample is warmed. The same procedure was followed using an annealed STO substrate with no film. The measurements from the STO substrate were subtracted from the measurements of the film to remove the contribution from the substrate. After this a small constant offset was subtracted from the FC and ZFC curves so that the magnetization is zero at 50K. A clear peak is seen in the ZFC data below T$_C$ at around 7 K, suggestive of a blocking temperature associated with superparamagnetism. This peak is also seen in thin ($\sim$10 QL thick) samples as well as in samples grown on InP(111)A substrates. Such peaks are typically associated with superparamagnetic or spin glass materials but have also been occasionally observed in ferromagnets. Further studies are needed to identify its origin.

For the field sweep, the sample is cooled in field to 5K and the applied field is swept from 9000 Oe to -9000 Oe and back. Again, the substrate contribution is subtracted. All measured magnetizations are normalized by the sample area. It should be noted that the features in the field sweep at +/-2500 Oe are artifacts due to the small magnetization signal in the raw data when the strong diamagnetic contribution of the substrate causes the total magnetization to cross through zero at these values. Using a composition of Cr$_{0.1}$(Bi$_{0.5}$Sb$_{0.5}$)$_{1.9}$Te$_3$ and a high field saturation magnetization of 7.25e-7 emu/mm$^2$ gives a moment of approximately 3.15 $\mu_B$/Cr, close the expected value of 3.

Given these results, a 5x5 $\mu m^2$ area of a 7QL sample of Cr$_{0.1}$(Bi$_{0.5}$Sb$_{0.5}$)$_{1.9}$Te$_3$ should show a magnetization change of $\sim$3.3$\times10^8\mu_B$ when integrating from zero magnetization to saturation value. This number is comparable to M obtained by cumulative addition of m in Fig. 2E. For 10QL sample the total change in M from negative to positive saturation should be $\sim$9.4$\times10^8\mu_B$ comparable to Fig. S13.

%%%%%%%%%%%%%%% FIGURE S14 %%%%%%%%%%%%%%%%%%%%%%%%%%%
\begin{figure}[H]
\includegraphics[width=\textwidth]{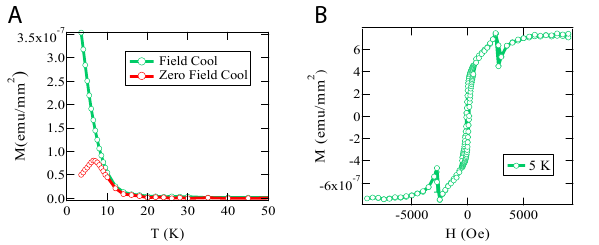}
\caption{\textbf{SQUID magnetometry measurements of a $\sim$40QL thick film.}
(A) M vs. T field cool and zero field cool magnetization curves showing a peak in the ZFC below T$_C$. 
(B) M vs. H hystersis curve at 5 K after subtracting the STO substrate contribution.  
}
\label{giantSQUID}
\end{figure}
%%%%%%%%%%%%%%%%%%%%%%%%%%%%%%%%%%%%%%%%%%%%%%%%%%%%%

\section{Transport and superparamagnetism in Mn-doped BiTe}

%%%%%%%%%%%%%%%% FIGURE S15 %%%%%%%%%%%%%%%%%%%%%%%%%%%
\begin{figure}[H]
\includegraphics[width=\textwidth]{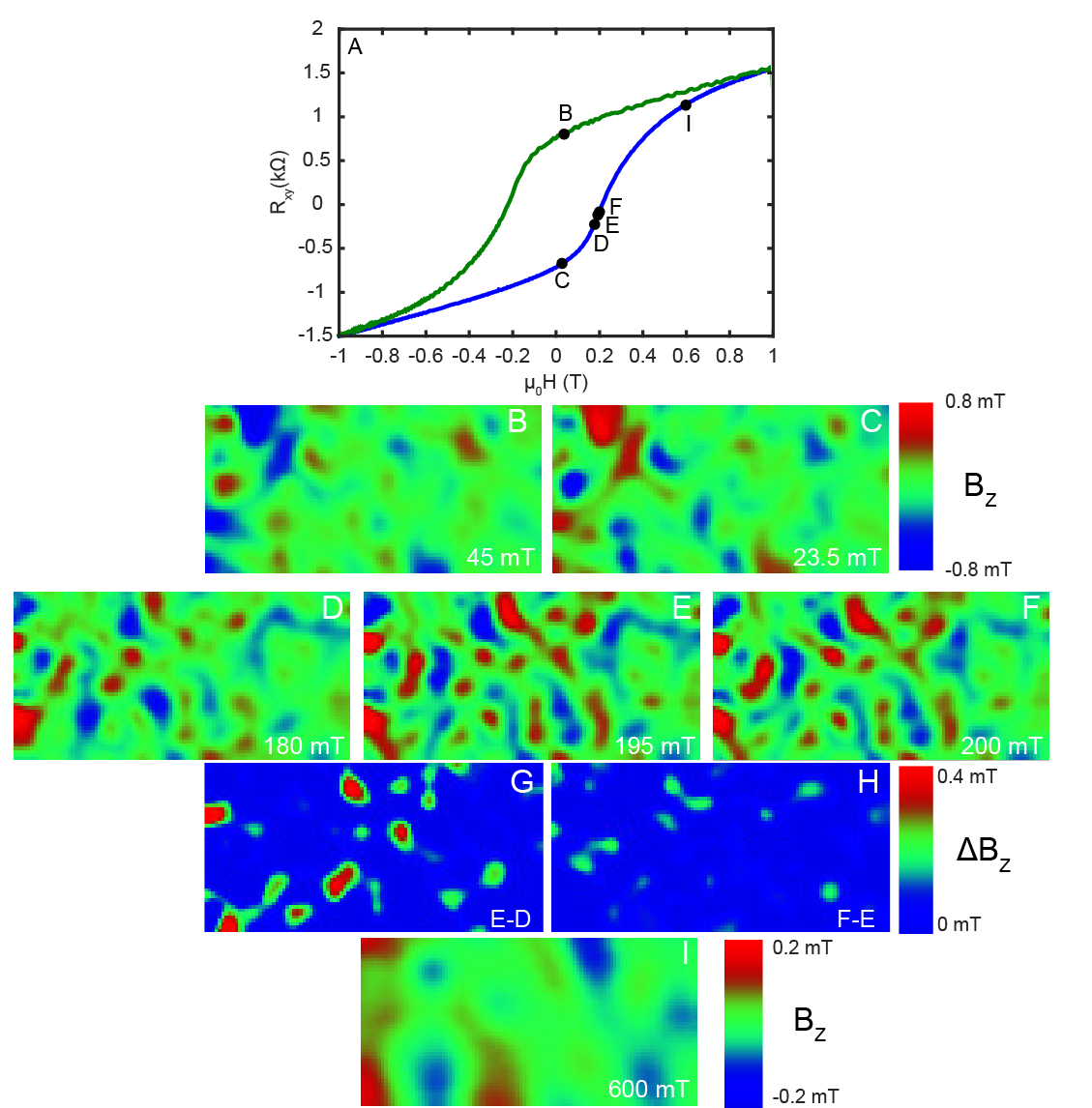}
\caption{\textbf{ Transport and scanning magnetic imaging of 70 nm thick Mn-doped BiTe sample at T=250 mK.}
(A) $R_{xy}$ vs. applied field showing hysteretic loop with coercive field of $H_c \simeq 0.2$ T in 70 nm thick Mn-doped BiTe sample grown by MBE on InP substrate. 
(B-F,I) Scanning SOT images $2.5\times 5$ $\mu$m$^2$ of the magnetic field $B_z(x,y)$ at six points along the magnetization loop marked in (A). Images (B) and (C) were acquired near zero applied field (45 mT and 23.5 mT respectively) on the opposite branches of the loop. The magnetic structure is highly anticorrelated similar to the behavior in Fig. 1.  Consecutive images at fields of 180 (D), 195 (E), and 200 mT (F) show superparamagnetic behavior with no domain wall motion. (G) Differential image $\Delta B_z(x,y)$ obtained by subtracting (D) from (E) showing isolated moment reversals. (H) Same as (G) for images (E) and (F).
(I) $B_z(x,y)$ at 0.6 T - nearly at the closing of the loop - demonstrating that though with lower contrast, the magnetic nano-structure prevails up to high fields. 
}
\label{MnSample}
\end{figure}
%%%%%%%%%%%%%%%%%%%%%%%%%%%%%%%%%%%%%%%%%%%%%%%%%%%%
\pagebreak
\section{Atomic Force Microscopy of Cr-doped samples}

%%%%%%%%%%%%%%%% FIGURE S13 %%%%%%%%%%%%%%%%%%%%%%%%%%%
\begin{figure}[H]
\includegraphics[width=\textwidth]{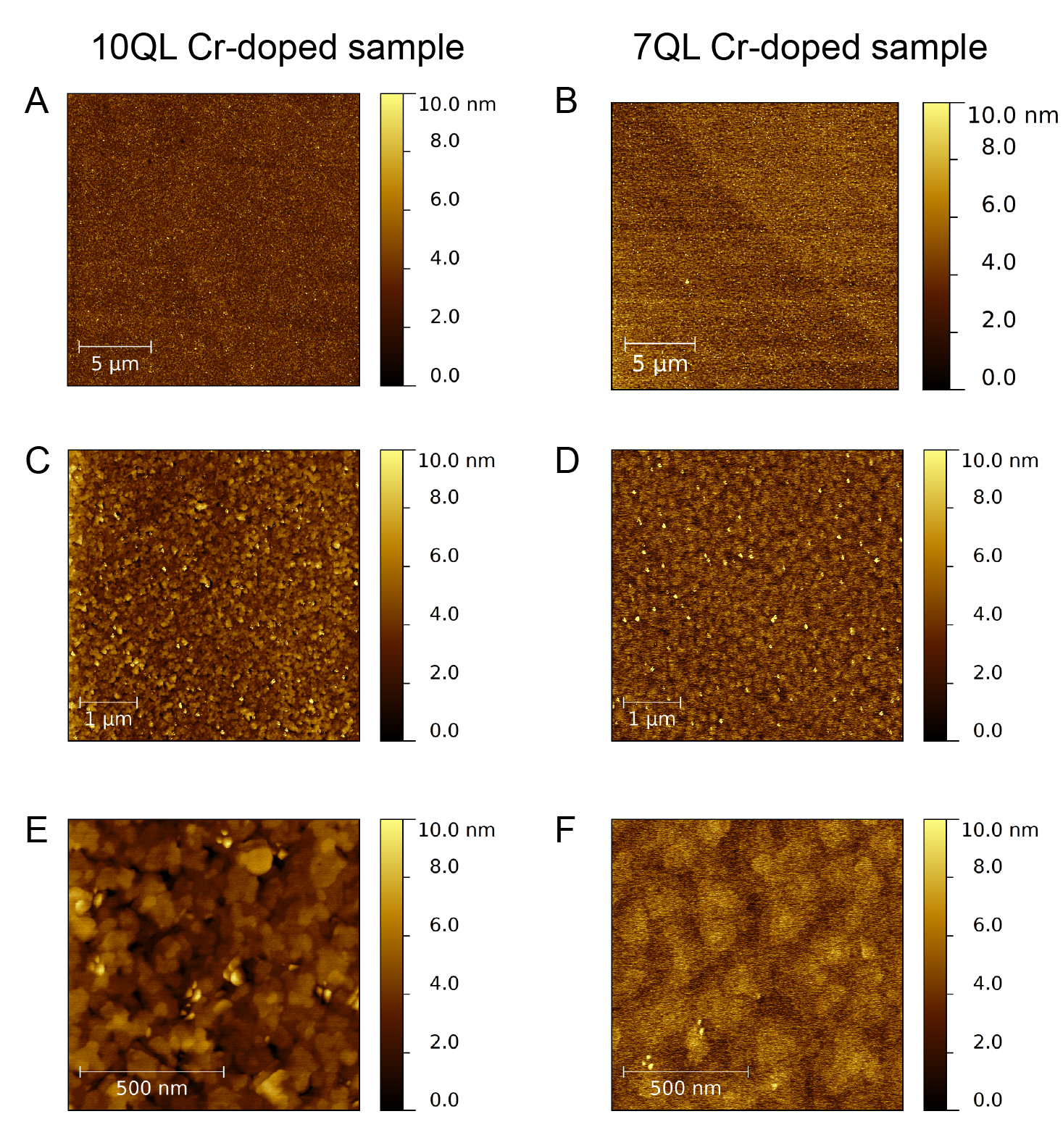}
\caption{\textbf{ AFM topography images of the two studied Cr-doped (Bi,Sb)$_2$Te$_3$ samples.} Both samples show surface disorder with characteristic lateral length scale of about 100 nm. X-ray diffraction shows that this surface roughness is created by the formation of a mosaic of twinned crystalline domains during epitaxial growth. The 7QL sample described in the main text (right column) has a more uniform morphology and lower surface roughness than the 10QL sample.
}
\label{AFM}
\end{figure}
%%%%%%%%%%%%%%%%%%%%%%%%%%%%%%%%%%%%%%%%%%%%%%%%%%%%